\documentclass[11pt,a4paper]{article}
\usepackage{jheppub}
\usepackage[curve]{xypic}
\usepackage[]{array}
\usepackage[]{amsmath}
\usepackage[]{amssymb}
\usepackage[]{mathrsfs}
\usepackage[]{tensor}
\usepackage[symbol]{footmisc}
\newcommand{\dd}{\partial}
\newcommand{\hdelta}{\hat{\delta}}

\newcommand{\tr}{\textrm{tr}}
\newcommand{\str}{\textrm{str}}
\newcommand{\qqd}{\quad , \quad}
\newcommand{\cc}{\mathbb{C}}

\newcommand{\be}{\begin{eqnarray}}
\newcommand{\ee}{\end{eqnarray}}
\newcommand{\0}{\nonumber}
\newcommand{\Gam}{\mathbf{\Gamma}}
\newcommand{\R}{\mathbf{R}}
\newcommand{\A}{\mathbf{A}}
\newcommand{\E}{\mathbf{E}}
\newcommand{\F}{\mathbf{F}}
\newcommand{\J}{\mathbf{J}}
\newcommand{\Q}{\mathbf{Q}}
\newcommand{\LL}{\mathbf{L}}
\newcommand{\Db}{\mathbf{D}}
\newcommand{\X}{\mathbf{\Xi}}
\newcommand{\Th}{\mathbf{\Theta}}
\newcommand{\CS}{\mathbf{\Upsilon}}
\newcommand{\nc}{^\mathrm{nc}}
\newcommand{\cov}{^\mathrm{cov}}
\newcommand{\Sig}{\mathbf{\Sigma}}
\newcommand{\bifeq}{\buildrel \mathcal{B} \over =}
\newcommand{\gwedge}{g}
\newcommand{\al}{\boldsymbol{ \alpha}}

\title{Gravitational Chern-Simons Lagrangians and black hole entropy\footnote{SISSA/89/2010/EP, \, YITP-10-113, \, ZTF-11-03}}

\author[a,d]{L.~Bonora,}
\author[b]{M.~Cvitan,}
\author[c]{P.~Dominis Prester,} 
\author[b]{S.~Pallua}
\author[a,b]{and I.~Smoli\'c}

\affiliation[a]{International School for Advanced Studies (SISSA/ISAS),\\
Via Bonomea 265, 34136 Trieste, Italy}
\affiliation[b]{Theoretical Physics Department, Faculty of Science, University of Zagreb,\\
p.p.~331, HR-10002 Zagreb, Croatia}
\affiliation[c]{Department of Physics, University of Rijeka, Omladinska 14, HR 51000, Croatia}
\affiliation[d]{Yukawa Institute for Theoretical Physics, Kyoto University, Kyoto 606-8502, Japan}

\emailAdd{bonora@sissa.it}
\emailAdd{mcvitan@phy.hr}
\emailAdd{pprester@phy.uniri.hr}
\emailAdd{pallua@phy.hr}
\emailAdd{ismolic@phy.hr}

\abstract{We analyze the problem of defining the black hole entropy when Chern-Simons terms are present in the action. Extending previous works, we define a general procedure, valid in any odd dimensions both for purely gravitational CS terms and for mixed gauge-gravitational ones.
The final formula is very similar to Wald's original formula valid for covariant actions, with a significant modification. Notwithstanding an apparent violation of covariance we argue
that the entropy formula is indeed covariant. 
}


\arxivnumber{\tt hep-th/yymm.xxxx}

\begin{document}


\maketitle

\section{Introduction}%
\label{sec:intro}

After some erratic appearances in the eighties, \cite{DJT1},\cite{DJT2} and \cite{Witten},
gravitational Chern-Simons (CS) action terms are met more and more frequently in the literature. This refers not only to a reappreciation of the earlier proposals (see, for instance, \cite{DT}, \cite{WittenM}) but, also, to more and more frequent appearances of CS terms as part of the gravity action in various dimensions. Low-energy effective actions coming from superstring theories provide many such examples. These terms are particularly interesting in relation to black holes. 
One of the problems they raise is how their addition modifies black hole solutions and associated charges. Another important problem is how to compute the black hole entropy in their presence. This is precisely what we wish to address in this paper.

While the 2+1 dimensional case can count by now on a considerable number of analyses
\cite{Solodukhin:2005ah,Perez:2010hk,Kraus:2005zm,Park:2006gt,Miskovic:2009kr}, 
little is known about higher dimensional gravitational Chern-Simons theories. 
In this paper we would like to start searching systematic answers to the questions raised by the presence of these terms in higher dimensions. We will extend our analysis also to reducible CS action terms containing non gravitational components.
 
To be less generic, let as introduce a few definitions and basic properties. In this paper we wish to investigate some properties of gravitational actions extended with Chern-Simons terms,
\be \label{lagrgen}
\mathcal{L} = \mathcal{L}_{\mathrm{cov}} + \mathcal{L}_{\mathrm{CS}}
\ee
By $\mathcal{L}_{\mathrm{cov}}$ we denote some generic manifestly diffeomorphism covariant gravitational Lagrangian in $D$ dimensions, while $\mathcal{L}_{\mathrm{CS}}$ contain Chern-Simons terms and, therefore, is not manifestly covariant. 

Formally, a purely gravitational Chern-Simons term in $D = 2n - 1$ dimensions is obtained by going to 
$(D+1)$ dimensions, where 
\be\label{dLCS}
P_n (\R, \dots, \R) = d \CS_{\mathrm{CS}}^{(n)} \;,
\ee
holds. Here $P_n$ denotes an invariant symmetric polynomial of a given Lie algebra (see Appendix A). 
We recall that for classical groups invariant polynomials can be defined via symmetric traces ($\str$) of the Lie algebra elements.
The $P_n$'s we have just introduced for purely gravitational CS terms are the invariant polynomials with respect to the Lie algebra 
of the $SO(2n-1)$ group. 
Eq.(\ref{dLCS}) can be explicitly integrated (transgression) \cite{chern},
\be\label{LCS}
\CS^{(n)}_{\mathrm{CS}}(\Gam) = n \int_0^1 dt \ P_n (\Gam, \R_t, \dots, \R_t)
\ee
The expression (\ref{LCS}) will be taken to define an action term in $D=2n-1$ dimensions, i.e.,
\be \label{LpCS}
\LL_{\mathrm{pCS}} = \CS^{(n)}_{\mathrm{CS}}
\ee
where subscript $\mathrm{pCS}$ denotes contribution of the pure gravitational CS term. We will often drop the superscript $^{(n)}$ whenever no confusion is possible.

It is important to recall that for $n=2k-1$, that is, in $D = 4k - 3$ dimensions, {\it irreducible} invariant symmetric polynomial vanish identically,
$P_{2k-1}= 0$.
So, purely gravitational irreducible CS terms can appear only in $D = 4k- 1$ dimensions. 
Generically $P_n$ will not be irreducible. 

We will also consider action terms where a gravitational CS term is multiplied by an invariant polynomial made of one or more gauge ``curvatures'', so as to fill up a $D$-form. Typically such reducible action terms will have the form
\be\label{mixCS}
\LL_{1,\mathrm{mix}} = \CS^{(m)}_{\mathrm{CS}}(\Gam)P_k(\F) \quad \textrm{or} \quad 
\LL_{2,\mathrm{mix}} = P_m(\R) \CS^{(k)}_{\mathrm{CS}}(\A)
\ee
where $\F$ represents the generic curvature of the gauge connection $\A$ and $P_k$ is an invariant polynomial of order $k$, such that $D= 2m+2k-1$. $\A$ can  be either a non-Abelian or an Abelian gauge connection, or even a RR field. We will refer to terms like (\ref{mixCS}) as {\it mixed Lagrangians}. Let us also notice that the two Lagrangian terms (\ref{mixCS}) descend via transgression from the same
term $P_m(\R)P_k(\F)$ in $(D+1)$ dimensions. We can have in general many of these fields simultaneously but, for simplicity, we will consider only one connection. The extension to several connections and curvatures is straightforward. Generally, one can have\footnote{It can be shown that Lagrangians in which CS terms appear inside the definition of gauge field strength, like $\mathbf{H} = d\mathbf{B} + \LL_{CS}$ in heterotic string theory, can be put in a classically equivalent form in which CS terms appear only through separate terms of the form (\ref{mixCS}), but not inside the definition of gauge field strengths. For derivation see, e.g., review \cite{Sen:2007qy}, and for applications in calculations in all orders in $\alpha'$ see \cite{Prester:2008iu,Prester:2009mc}.}
\be \label{LCSgen}
\LL_{\mathrm{CS}} = \sum_i \LL_{\mathrm{gCS}}^{(i)} + \sum_j \LL_{\mathrm{aCS}}^{(j)}
\ee
where
\be
\LL_{\mathrm{gCS}}^{(i)} &=& \CS^{(m_i)}_{\mathrm{CS}}(\Gam) \, P_{n_i}(\R) \,
  P_{k_{i1}}(\F_1) \, P_{k_{i2}}(\F_2) \cdots
\0 \\
\LL_{\mathrm{aCS}}^{(j)} &=& \CS^{(m_j)}_{\mathrm{CS}}(\A_{j}) \, P_{n_j}(\R) \,
  P_{k_{j1}}(\F_1) \, P_{k_{j2}}(\F_2) \cdots \0
\ee

In our subsequent calculations the various independent pieces of the Lagrangian yield independent contributions to the intermediate and final results. As a consequence they can be dealt with separately (this is generically not true for the equations of motion, but, as we will see, the latter need not be explicitly calculated). Therefore to make the presentation more clear, we will mostly use pure gravitational CS term $\LL_{\mathrm{pCS}}$ to demonstrate the procedures. Calculations for the mixed CS terms usually proceed in a similar fashion. The total contribution to the results are obtained by summing the contributions of all the separate terms appearing in the total Lagrangian.

A few comments on general CS terms are perhaps useful.
They were introduced in the mathematical literature by S.S.~Chern and J.~Simons in 1974 (see \cite{CS}) to represent characteristic classes of the total space of a principal bundle. Gravitational CS terms have been introduced in theoretical physics mostly to describe topological theories. One should not forget that in $\LL_{\mathrm{CS}}(\Gam)$, as well as in (\ref{mixCS}), there is no explicit dependence on the metric --- the metric appears only through $\Gam$. It is worth recalling that, while general gauge CS terms are not globally defined on the base space of a principal bundle (because a gauge connection $\A$ is generally not basic), the gravitational CS terms are globally defined in space-time due to the existence of a natural lift in the frame bundle, see for instance \cite{BCRS}. 

After these somewhat lengthy preliminaries let us come to the subject of the present paper, where
we are interested in studying some consequences of introducing in a theory additional CS terms, in particular in the modifications induced in asymptotic or boundary charges, with emphasis on black hole entropy. For manifestly covariant theories with covariant Lagrangians, there is a powerful method based on the  covariant phase space formalism, which was adopted by Wald \cite{IyerWald} and led to an elegant general formula for black hole entropy. In \cite{Tach} the author outlined an extension of this elegant formalism to theories which are not manifestly covariant (Lagrangians are not covariant) but have covariant equations of motion, such as (\ref{lagrgen}), and applied it to the calculation of the black hole entropy in some instances when a CS term is present. In this paper we wish to carry this program to completion, so as to be able to apply it in general. More precisely
we will give a more detailed and elaborate analysis of the general method outlined in \cite{Tach} and apply it to general theories with CS terms in the Lagrangian (\ref{lagrgen}). In the process we will introduce some
necessary modifications with respect to \cite{Tach} in order to guarantee consistency. We will compute in detail the CS induced modifications to Wald entropy formula, and find out that the final formula is very similar to Wald's original formula valid for covariant actions, with a significant modification. We will analyze the subtleties connected to covariance and to the use of different types of coordinate systems. Although the modified Wald formula for entropy
looks non-covariant, we will show that it can be `covariantized'.

The paper is organized as follows. In the next section we introduce the main formulas and derive the generalized Cotton tensor and the symplectic potential. In section 3 we generalize the covariant phase space formalism to the case of actions containing CS terms. In section 4 we derive the entropy formula and discuss its covariance. Section 5 is for the conclusions.
To render the paper more readable many of the calculations are postponed to dedicated appendices.

\section{Equations of motion}%
\label{sec:eom}

Let us consider first the pure gravitational CS term in the Lagrangian (\ref{LpCS})-(\ref{LCS}). In order to compute contribution to the equations of motion we calculate the (generic) variation with respect to 
$\Gam$. Using
\be
\delta \Gam_t = t \delta \Gam \qqd \delta \R_t = d \delta \Gam_t + [\Gam_t, \delta \Gam_t]
 = \Db_t \delta \Gam_t
\ee
we have
\be
\delta \LL_{\mathrm{pCS}}(\Gam) = n \int_0^1 dt \left( P_n (\delta \Gam, \R_t, \dots, \R_t)
 + (n-1) P_n (\Gam, \Db_t \delta \Gam_t, \R_t, \dots, \R_t) \right)
\ee
Using (\ref{dPn}) this can be put in the following form,
\be
\delta \LL_{\mathrm{pCS}}(\Gam) &=& n \int_0^1 dt \, \Big( P_n (\delta \Gam, \R_t, \dots, \R_t)
 - (n-1) d P_n (\Gam, \delta \Gam_t, \R_t, \dots, \R_t) +\0\\
&& + (n-1) P_n (\Db_t \Gam, \delta \Gam_t, \R_t, \dots, \R_t) \Big)
\ee
It is easy to see that
$$(n-1) P_n (\Db_t \Gam, \delta \Gam_t, \R_t, \dots, \R_t)
 = (n-1) P_n (\dd_t \R_t, \delta \Gam_t, \R_t, \dots, \R_t) =$$
\be
= \frac{d}{dt} P_n(\delta \Gam_t, \R_t, \dots, \R_t) - P_n(\delta \Gam, \R_t, \dots, \R_t)\0
\ee
This allows us to directly integrate one of the terms, obtaining finally
\be
\delta \LL_{\mathrm{pCS}}(\Gam) = n \left( P_n(\delta \Gam, \R, \dots, \R)
 - (n-1) \, d \int_0^1 dt \ P_n (\Gam, \delta \Gam_t, \R_t, \dots, \R_t) \right)\label{deltaLCS1}
\ee
By adopting a compact notation we can write this as
\be\label{deltaLCS2}
\delta \LL_{\mathrm{pCS}}(\Gam) = n \, P_n (\delta \Gam \, \R^{n-1}) + d\Th\nc(\Gam, \delta\Gam) 
\ee
where
\be\label{Thetanc}
\Th\nc \, \equiv \, - n(n-1) \, \int_0^1 dt \ P_n (\Gam, \delta \Gam_t, \R_t^{n-2})
\ee

It is immediate to generalize (\ref{deltaLCS1}, \ref{deltaLCS2}) to mixed Lagrangians,
\be
\delta\LL_{1,mix} &=& m P_m(\delta \Gam, \R^{m-1}) P_k(\F) + k P_m(\R) P_k(\delta \A, \F^{k-1}) - \label{mixed1}\\
&&- d \, \Bigl( m(m-1) \int_0^1 dt \, P_m(\Gam, \delta\Gam_t, \R_t^{m-2}) P_k(\F)+ k \CS_{\mathrm{CS}}^{(m)}(\Gam) P_k(\delta\A,\F^{k-1})\Bigr)\0
\ee
\be
\delta\LL_{2,mix} &=& m P_m(\delta \Gam, \R^{m-1}) P_k(\F) + k P_m(\R) P_k(\delta \A, \F^{k-1}) + \label{mixed2}\\
&&+ d \, \Bigl( mP_m(\delta\Gam, \R^{m-1}) \CS_{\mathrm{CS}}^{(k)}(\A) - k(k-1)P_m(\R) \int_0^1 dt P_k(\A, \delta \A_t, \F_t^{k-2}) \Bigr)\0
\ee

Let us concentrate first on (\ref{deltaLCS1}). We introduce the $2(n-1)$-form $\mathbf{K} \equiv \R^{n-1}$; its components are
\be
\tensor{K}{^\alpha^\beta_{\mu_1}_{\mu_2}_\cdots_{\mu_{2n-2}}} = \frac{(2n-2)!}{2^{n-1}} \, \tensor{R}{^\alpha_{\sigma_1}_[_{\mu_1}_{\mu_2}} \, \tensor{R}{^{\sigma_1}_{| \sigma_2 |}_{\mu_3}_{\mu_4}} \cdots \tensor{R}{^{\sigma_{n-2}}^{\beta}_{\mu_{2n-3}}_{\mu_{2n-2}]}}
\ee
This means that we have
\be
\delta \LL_{\mathrm{pCS}} = n(2n-1) \tensor{K}{^\alpha_\lambda_[_{\mu_1}_{\mu_2}_\cdots_{\mu_{2n-2}}} \delta \Gamma^\lambda_{|\alpha|\mu_{2n-1}]} \, dx^{\mu_1} \cdots dx^{\mu_{2n-1}} + d\Th\nc
\ee
Using the Levi-Civita tensor the previous formula can be rewritten as
\be
\delta \LL_{\mathrm{pCS}} = \frac{n(-1)^s}{(2n-2)!} \, \epsilon^{\mu_1 \cdots \mu_{2n-1}}
 \tensor{K}{^\alpha_\lambda_{\mu_1}_{\mu_2}_\cdots_{\mu_{2n-2}}}
 \delta \Gamma^\lambda_{\alpha \mu_{2n-1}} \, \sqrt{-g} \, d^{2n-1} x + d\Th\nc
\ee
where $s$ denotes the metric signature. Now, using 
\be
\delta \Gamma^\lambda_{\alpha \mu} = \frac{1}{2} \, g^{\lambda \beta} \left( \nabla_{\!\alpha} \, \delta g_{\beta \mu} + \nabla_{\!\mu} \, \delta g_{\beta \alpha} - \nabla_{\!\beta} \, \delta g_{\alpha \mu} \right)
\ee
and Bianchi identity we obtain 
\be \label{dLpCS}
\delta \LL_{\mathrm{pCS}} = C^{\beta \mu} \delta g{}_{\beta \mu} \,  \boldsymbol{\epsilon} 
 + d\Th\cov + d\Th\nc
\ee
where $\boldsymbol{\epsilon} = \sqrt{-g} \, d^{2n-1} x$ is the volume $D$-form, the components of 
$\Th\cov$ are
\be
{\Theta\cov}\tensor{}{_{\mu_1}_{\mu_2}_\cdots_{\mu_{2n-2}}}
&=& (-1)^s n \,  \tensor{(*K)}{^\alpha^\beta^\mu} \delta g{}_{\beta \mu} \,  \tensor{\epsilon}{_\alpha_{\mu_1}_{\mu_2}_\cdots_{\mu_{2n-2}}} \nonumber\\
&=& n(2n-1) \tensor{K}{^\alpha^\beta_[_{\mu_1}_{\mu_2}_\cdots_{\mu_{2n-2}}}  \delta^\mu_\alpha{}_{]} \delta g{}_{\beta \mu} \nonumber\\
&=& -2 n^{2} \tensor{ \delta g}{^{\vphantom{\beta}}_\beta_[_{\mu_1}} \tensor{K}{^\alpha^\beta_{|\alpha|}_{\mu_2}_\cdots_{\mu_{2n-2}}_]}  \quad \textrm{(even $n$) }
\label{Thetacov}
\ee
and
\be
C^{\beta \mu}  = - (-1)^s n \,   \nabla{}_\alpha \tensor{(*K)}{^\alpha^{(\beta}^{\mu)}} 
\ee
where $*\mathbf{K}$  is the Hodge dual of $\mathbf{K}$: 
\be
\tensor{(*K)}{^\alpha^\beta^\mu} \equiv \frac{1}{(2n-2)!} \, 
\tensor{K}{^\alpha^\beta_{\mu_1}_{\mu_2}_\cdots_{\mu_{2n-2}}}
\tensor{\epsilon}{^{\mu_1}^{\mu_2}^\cdots^{\mu_{2n-2}}^\mu} \, 
\ee
Obviously $C^{\beta \mu}$ is the contribution to the equations of motion of the $\LL_{\mathrm{pCS}}$ term.
It is a symmetric, traceless and divergence free (pseudo)tensor, investigated in some detail in 
\cite{Solodukhin:2005ns} where it was shown that it can be viewed as a generalization of the Cotton tensor to $D=2k+1$, $k>1$ dimensions \cite{Solodukhin:2005ns,Garcia}.

As can be read off from (\ref{dLpCS}), the total \emph{symplectic potential} $\Th$ is the sum of the two terms
$\Th\cov$ (given by (\ref{Thetacov})) and $\Th\nc$ (given by (\ref{Thetanc})). We refer to $\Th\cov$ and $\Th\nc$ 
as the \emph{covariant} and the \emph{noncovariant} contribution to the symplectic potential, respectively.

\section{Covariant phase space formalism for noncovariant Lagrangians}%
\label{sec:covphsp}

The covariant phase space formalism is a powerful tool to calculate various charges connected with 
asymptotic  symmetries. For example, it was used in \cite{Wald1,JKM,IyerWald} for an elegant derivation 
of the first law of black hole thermodynamics in theories with manifestly diffeomorphism covariant Lagrangian descriptions. In \cite{Tach} the formal extension to theories with Lagrangian description which is not manifestly diff-covariant was outlined. In this section we wish to apply this construction to theories with general Chern-Simons terms. In the process we calculate all the necessary ingredients that will be used in the next section to obtain a general formula for black hole entropy in such theories, extending in this way the results of \cite{IyerWald,Tach}.   

In what follows, as usual, we understand that the Lagrangian is a $D$-form. Also we denote by the symbol ``$\phi$'' all the dynamical fields, that is the spacetime metric $g_{ab}$ as well as any matter field. The derivation of the first law in \cite{Wald1} assumes a diff-covariant Lagrangian, for which the following holds,
\be
\delta_\xi \LL_{\mathrm{cov}}(\phi) = \pounds_\xi \LL_{\mathrm{cov}}(\phi)\0
\ee
where $\delta_\xi$ is the variation induced by the diffeomorphism generated by the vector field $\xi$ with components $\xi^a$, while $\pounds_\xi$ denotes the Lie derivative with respect to $\xi$. This condition is \emph{not} satisfied if there are Chern-Simons terms in the Lagrangian; instead, we have
\be\label{deltaxiCS}
\delta_\xi \LL(\phi) = \pounds_\xi \LL(\phi) + d \X_\xi
\ee 
for some $(D-1)$-form $\X_\xi$. The form of (\ref{deltaxiCS}) expresses the fact that the action is
still diff-invariant (if the space-time is closed), though the Lagrangian is not diff-covariant. It is understood that
$\pounds_\xi$ acts on non-tensor quantities (such as $\Gam$) as if their indices were tensorial ones. As already pointed out above, in 
\cite{Tach} the  problem of extending the covariant phase space procedure to Chern-Simons type Lagrangians was first dealt with. We will follow in part this recipe and introduce the necessary modifications.

The first-order variation of the Lagrangian $\LL$ takes the general form
\be
\delta \LL (\phi) = \mathbf{E} \, \delta \phi + d \Th(\phi, \delta\phi)
\ee
where the $D$-form $\mathbf{E}$ represents the equation of motion\footnote{Summation over the dynamical fields and contraction of their tensor indices with corresponding dual tensor indices of $\mathbf{E}$ are understood.} (which is manifestly diff-covariant) and the `surface term', 
$(D-1)$-form $\Th$, is known as \emph{symplectic potential}, as we have already pointed out. In Sec. \ref{sec:eom} we have already found that CS terms in 
the Lagrangian introduce additional contribution to the symplectic potential which can be separated in 
covariant and noncovariant part
\be
\Th = \Th\nc + \Th\cov
\ee
where in the case of pure gravitational CS Lagrangian term (\ref{LpCS}), $\Th\nc$ and $\Th\cov$ are 
given by (\ref{Thetanc}) and (\ref{Thetacov}), respectively.

For the two mixed CS terms (\ref{mixCS}) we obtain  
\be
\Th_1\nc(\Gam,\delta\Gam, \A,\delta \A) &=&
 - m(m-1) \int_0^1 dt \, P_n(\Gam, \delta\Gam, \R_t^{m-2})P_k(\F) - \0\\
&-& k \CS^{(m)}(\Gam) P_k(\delta\A,\F^{k-1}) \label{ThetaGammaA1}
\ee
and
\be\label{ThetaGammaA2}
\Th_2\nc(\Gam,\delta\Gam, \A,\delta \A) &=& mP_m(\delta\Gam,\R^{m-1})\CS^{(k)}(\A) - \0\\
&-& k(k-1)P_m(\R) \int_0^1 dt \, P_k(\A, \delta\A_t, \F_t^{k-2}),
\ee
respectively. The covariant parts can be easily obtained, but, as we shall see, they are not essential for deriving the black hole entropy and so we shall not bother to write them
down.

Next, we define the non-covariant part of the diff transformation by
\begin{equation}
\hdelta_\xi \, \equiv \, \delta_\xi - \pounds_\xi
\end{equation}
Let us recall that for tensor-valued $p$-forms (like 1-form $\Gam$ and 2-form $\R$) the Lie-derivative satisfies
\begin{equation} \label{liedi}
\pounds_\xi = d \, \imath_\xi + \imath_\xi \, d + [\Lambda,\;\,] \;,\quad \Lambda{}^b{}_a= \partial_a\xi^b.
\end{equation}
In particular,
\be
\delta_{\xi}\Gam = \pounds_{\xi} \Gam + \hdelta_\xi \Gam \qqd \hdelta_\xi \Gam = d\Lambda
\label{deltaGamma}
\ee
This allows us to define the $\X_{\xi}$ term via the relation
\be
\hdelta_\xi \LL(\Gam) = d \X_\xi(\Gam)
\ee
For the pure gravitational Chern-Simons Lagrangian (\ref{LpCS}) we get
\be\label{XiGamma}
\X_\xi^{(n)} (\Gam) = n(n-1) \int_0^1 dt \, (t-1) P_n (d\Lambda, \Gam, \R_t^{n-2})
\ee
For the mixed Lagrangian case we have simply
\be\label{XiGammaA}
\X_{1,\xi} (\Gam,\A) = \X^{(m)}_\xi(\Gam) P_k(\F) \quad \textrm{and} \quad \X_{2,\xi} (\Gam,\A) = 0
\ee

One basic ingredient we need is the Noether charge $\Q_{\xi}$. This is defined
via the Noether current $(D-1)$-form $\J_\xi$, 
\be \label{Jdef}
\J_\xi = \Th_\xi - \imath_\xi \LL - \X_\xi
\ee
which is conserved on-shell (that is, for the field configurations that satisfy $\mathbf{E} = 0$), 
$$d\J_\xi = d\Th_\xi - d(\imath_\xi \LL) - d\X_\xi \approx \delta_\xi \LL - d(\imath_\xi \LL) - d\X_\xi = \pounds_\xi \LL - d(\imath_\xi \LL) = 0.$$
where $\approx$ denotes equations that are valid on-shell.
Here we have also used (\ref{liedi}), which for a Lagrangian $\LL$ reduces to
\be
\pounds_\xi = d \, \imath_\xi + \imath_\xi \, d
\ee
and the fact that $d \LL = 0$ since it is $D$-form. This implies, according to \cite{Wald1,Wald2}, that $\J_\xi$ is exact 
on-shell, i.e.~there is $(D-2)$-form $\Q_\xi$, such that
\be
\J_\xi \approx d\Q_\xi \label{J=dQ}
\ee
We refer to this property as {\it Wald lemma}.

To find $\Q_\xi$, as a local expression of the dynamical fields linear in $\xi$, one can use either the constructive method from \cite{Wald2}, or more formal
methods based on cohomology and the so-called variational complex. In all our cases $\Q_\xi$ takes the form
\be \label{Q10}
\Q_\xi = \Q_\xi^{(1)} + \Q_\xi^{(0)}
\ee
where the superscript denotes the number of derivatives acting on $\xi$. The details of calculations are presented in Appendix \ref{app:Qcalc}.

The results are as follows. For the purely gravitational case we obtain for $\Q_\xi^{(1)}$
\be \label{Qxi1}
\Q_{\xi}^{(1)}(\Gam) = n(n-1) \int_0^1dt P_n(\Lambda,\Gam,\R_t^{n-2})
\ee
and for $\Q_\xi^{(0)}$ (in components)
\be 
(Q_\xi^{(0)})_{\mu_1\cdots\mu_{2n-3}}
 &=& n(n-1) \int_0^1 dt P_n(i_\xi \Gam,\Gam_t,\R_t^{n-2})_{\mu_1\cdots\mu_{2n-3}} \0 \\
&& - n \, \xi_b \, K{}^{ab}{}_{a\mu_1\cdots\mu_{2n-3}}
 + \frac{n}{2}(2n-1) \xi_{[a} K{}^{ab}_{b\mu_1\cdots\mu_{2n-3}]}
\label{Q2new}
\ee

For the first mixed CS term we have
\be \label{Q1xiA}
\Q_{1,\xi}^{(1)}(\Gam,\A) =  m(m-1) \int_0^1dt P_m(\Lambda,\Gam,\R_t^{m-2})P_k(\F)
\ee
and for the second: 
\be
\Q_{2,\xi}^{(1)}(\Gam,\A) &=&  m P_m(\Lambda,\R^{m-1}) \CS_{\mathrm{CS}}^{(k)}(\A)\label{Q2xiA}
\ee
As we will explain later, only the $\Q_{\xi}^{(1)}$ part is relevant to the construction of the black hole entropy. 

Proceeding further we can calculate $\delta \delta_\xi \LL$ in two ways,
$$\delta \delta_\xi \LL = \delta \pounds_\xi \LL + \delta d \X_\xi = \delta d (\imath_\xi \LL) + \delta d \X_\xi \approx d \imath_\xi d \Th + \delta d \X_\xi$$
$$\delta_\xi \delta \LL \approx \delta_\xi d \Th = d \delta_\xi \Th = d (\pounds_\xi \Th) + d \hdelta_\xi \Th = d \imath_\xi d \Th + d \hdelta_\xi \Th$$
By comparison we see that
\be \label{dhTddelX}
d \hdelta_\xi\Th \approx \delta d\X_\xi = d \delta \X_\xi
\ee
having used $[\delta,d]=0$ in the last equality. This allows us to introduce a $(D-2)$-form $\mathbf{\Sigma}_\xi$ via
\be\label{defSigma}
\hdelta_\xi \Th - \delta \X_\xi \approx d\mathbf{\Sigma}_\xi
\ee
With the usual methods, for the purely gravitational CS Lagrangians, one can compute (see Appendix B.2)
\be\label{dSigma}
\mathbf{\Sigma}_\xi(\Lambda, \Gam) = -n(n-1)(n-2) \int_0^1 dt \, t(t-1) P_n (d\Lambda, \Gam, \delta \Gam, \R_t^{n-3})
\ee
We remark that, actually, in this case we have an exact relation:
\be\label{defSigmaexact}
\hdelta_\xi \Th - \delta \X_\xi = d\mathbf{\Sigma}_\xi
\ee

In a similar way, for the first mixed Lagrangian we have the exact relation
\be\0
\hat \delta_{\xi}\Th_1^{nc} (\Gamma,\A)-\delta \Xi_{1,\xi}(\Gamma,\A)= d\Sig_{1,\xi}(\Gam,\A)
\ee
where
\be
\mathbf{\Sigma}_{1,\xi}(\Gamma,\A)&=& -m(m-1)(m-2) \int_0^1 dt (t^2-t) P_m(d\Lambda, \Gam, \delta\Gam, \R_t^{m-3}) P_k(\F)\0\\
&& - km(m-1) \int_0^1 dt (t-1) P_m(d\Lambda,\Gam, \R_t^{m-2}) \, P_k(\delta \A ,\F^{k-1})\label{SigmaGammaA}
\ee
Finally, it is easy to prove that $\Sig_{2,\xi}\equiv 0$.

A central element in covariant phase space approach is the symplectic current $(D-1)$-form 
$\boldsymbol{\omega}$ defined by
\be \label{omth}
\boldsymbol{\omega}(\phi, \delta_1 \phi, \delta_2 \phi) = \delta_1 \Th(\phi, \delta_2 \phi)
 - \delta_2 \Th(\phi, \delta_1 \phi)
\ee
Let $\mathcal{C}$ be a Cauchy surface with orientation given by 
$$\hat{\boldsymbol{\epsilon}}_{a_1 \cdots \, a_{\mathrm{D-1}}}
 = n^b \boldsymbol{\epsilon}_{ba_1 \cdots \, a_{\mathrm{D-1}}}$$
where $n^a$ is the future pointing normal to $\mathcal{C}$ and $\boldsymbol{\epsilon}$ is the positively oriented spacetime volume form. We define the presymplectic form $\Omega$, a 2-form in the space of fields and 0-form in spacetime, by integrating the $(D-1)$-form $\boldsymbol{\omega}$ over a Cauchy surface 
$\mathcal{C}$,
\be \label{omega}
\Omega(\phi, \delta_1 \phi, \delta_2 \phi) = 
\int_\mathcal{C} \, \boldsymbol{\omega}(\phi, \delta_1 \phi, \delta_2 \phi)
\ee
The Hamilton equations of motion generated by vector field $\xi$ are then
\be \label{homega}
\delta H[\xi] = \Omega(\phi, \delta \phi, \delta_\xi \phi)
\ee
where $H[\xi]$ is Hamiltonian for $\xi$. From (\ref{omth}) and (\ref{Jdef}) one obtains (see \cite{Tach})
\be \label{dJomega}
\delta \J_\xi \approx \boldsymbol{\omega}(\phi, \delta \phi, \delta_\xi \phi)
 + d(\imath_\xi \Th + \mathbf{\Sigma}_\xi)
\ee
which, by using (\ref{omega}-\ref{dJomega}), leads to
\be \label{dHdQ}
\delta H[\xi] \approx \int_\mathcal{\partial C} \left( \delta \Q_\xi - \imath_\xi \Th - \mathbf{\Sigma}_\xi \right)
\ee
We see that the existence of $H[\xi]$ requires either that the last two terms on the right hand side of 
Eq. (\ref{dHdQ}) give vanishing contribution, or that there is a $(D-2)$-form $\mathbf{C}_\xi$ satisfying 
\be \label{Cint}
\delta \mathbf{C}_\xi = \imath_\xi \Th + \mathbf{\Sigma}_\xi
\ee
If the latter were the case one could introduce $\Q'_\xi = \Q_\xi - \mathbf{C}_\xi$ and  integrate Eq. (\ref{dHdQ}) to obtain 
\be \label{HQprime}
H[\xi] \approx \int_\mathcal{\partial C} \Q'_\xi
\ee
However, even for the pure gravitational Chern-Simons Lagrangian of the form (\ref{LpCS}) it seems to be generically very difficult to construct a $\mathbf{C}_\xi$ satisfying (\ref{Cint}) off-shell.
We therefore turn to the first alternative, which requires studying the problem on a case by case basis \cite{Wald:1999wa}. 

In our application to black hole entropy our strategy will be the following one:
first perform the calculations in Kruskal-type coordinates, then show that $ \imath_\xi \Th + \mathbf{\Sigma}_\xi$ vanishes, and finally ``covariantize" the results.

\section{Black hole entropy in theories with Chern-Simons terms}
\label{sec:CSentropy}

\subsection{BH Thermodynamics in Kruskal coordinates}
\label{ssec:entKrus}

Wald and others (\cite{Wald1},\cite{JKM},\cite{IyerWald}) have shown how to compute the black hole entropy using a covariant phase space formalism. They have considered a general classical theory of gravity in $D$ dimensions, arising from a diffeomorphism-invariant Lagrangian. Assuming that the theory admits a stationary black hole solution with a bifurcate Killing horizon and that the canonical mass and angular momentum of the solution are well defined at infinity, the first law of black hole mechanics always holds for perturbations of nearly stationary black hole solutions. Using this, it was shown (\cite{IyerWald}) that for a given Lagrangian $\mathcal{L}$, the entropy of a black hole can be computed by an elegant formula
\be \label{waldent}
S = -2\pi \int_\Sigma d^{D-2}x \, \sqrt{-g} \, \frac{\delta \mathcal{L}}{\delta R_{abcd}} \, 
 \epsilon_{ab} \epsilon_{cd}
\ee
where the integration is over the horizon cross section $\Sigma$ with binormal 
$\epsilon_{ab}$ normalized as $\epsilon_{ab}\epsilon^{ab} = -2$. We now want to find a generalization 
of Wald entropy formula (\ref{waldent}) to the theories with Chern-Simons terms of the most general form.

Following \cite{IyerWald,Tach} we assume $\xi$ to be a Killing vector field, and that dynamical fields are symmetric, i.e., $\delta_\xi \phi = 0$. Then (\ref{dHdQ}) gives 
\be \label{dQKill}
\int_\mathcal{\partial C} \left( \delta \Q_\xi - \imath_\xi \Th - \mathbf{\Sigma}_\xi \right)
\approx \delta H[\xi] \approx  0
\ee
For an isolated black hole the boundary of the Cauchy surface $\mathcal{\partial C}$ consists of two disconnected
parts, one is an intersection of $\mathcal{C}$ with the horizon, which we denote by $\Sigma$, and the other is the asymptotic infinity (denoted by $\infty$). So (\ref{dQKill}) becomes
\be \label{dQbh}
\int_\Sigma \left( \delta \Q_\xi - \imath_\xi \Th - \mathbf{\Sigma}_\xi \right)
 \approx \int_\infty \left( \delta \Q_\xi - \imath_\xi \Th - \mathbf{\Sigma}_\xi \right)
\ee
As in \cite{IyerWald} we assume the black hole has Killing horizon, generated by $\xi$, and that the surface
gravity $\kappa$ is constant on the horizon.\footnote{It was shown in \cite{Racz:1995nh} that 
 $\kappa$ is constant on Killing horizon for all stationary-axisymmetric black holes with the "$t-\varphi$" reflection isometry.} One consequence is that such spacetime admits an extension in which the horizon has a bifurcation surface $\mathcal{B}$ \cite{Racz:1992bp}, on which
\be \label{xibf0}
\xi \big|_\mathcal{B} = 0
\ee
This property makes it extremely useful to choose the Cauchy surface such that  $\Sigma = \mathcal{B}$. From now on we are assuming this.

Now we want to show that (\ref{dQbh}) yields the first law of thermodynamics, namely
\be \label{TdS}
T \delta S = \delta U - \ldots \;.
\ee

The left hand side of (\ref{dQbh}) needs a more careful analysis, especially as the individual terms are
(at least apparently) not covariant. In ``regular" coordinate systems (which are well-defined at the
bifurcation surface $\mathcal{B}$), property (\ref{xibf0}) implies that the second term in (\ref{dQbh})
vanishes
\be \label{ixiTHB}
\imath_\xi \Th \big|_\mathcal{B} = 0 \;,
\ee
From the same argument it follows that we can drop the $\Q_\xi^{(0)}$ part of $\Q_\xi$. Also, in coordinate systems regular at $\mathcal{B}$, we can write $\Lambda$ in 
the following covariant way
\be \label{lambf}
\Lambda^a_{\;\; b} \big|_\mathcal{B} = \nabla_b \xi^a \big|_\mathcal{B} = \kappa \, \epsilon^a_{\;\; b}  
\ee
where $\epsilon$ is the binormal of $\mathcal{B}$.

We still have to deal with the second term in (\ref{dQbh}). As we do not know how to generically 
integrate $\mathbf{\Sigma}_\xi$,
we turn to a different strategy. From the results of the previous section we saw that $\mathbf{\Sigma}_\xi$ contains a $d\Lambda$ factor. Consequently it appears beneficial to specialize for the moment to a particular ``Kruskal-type" coordinate system (details are in Appendix 
\ref{app:Kruskal}) in which
\be
d\Lambda = 0 \;.
\ee
This makes the $\mathbf{\Sigma}_\xi$ term trivially vanish\footnote{What is at stake in relation to the $\Sig_{\xi}$ term (which is anyhow absent in  3D) is the following. If
$\Sig_{\xi}$ is not integrable, i.e. there does not exists any local ${\bf C}_\xi$ such that 
$\Sig_{\xi}=\delta {\bf C}_{\xi}$, then a charge $\Q_\xi$
can still be formally obtained by integrating $\Sig_\xi$ along a path in
the space of solutions, starting from some conventional one. However
in this way the result turns out to be path dependent. Path dependence of
this result and  $\Sig_\xi$ non-integrability are two equivalent
statements.  In this situation the entropy would not be uniquely defined.
It follows that we are obliged to assume $\Sig_\xi$--integrability.
However, after assuming it, we still do not know the explicit form of
${\bf C}_\xi$, which is very hard to construct. The procedure described in this section teaches us
how to avoid such explicit construction.
}.

From the form of $\Q_\xi^{(1)}$, obtained in the previous section, and (\ref{lambf}), it follows that the left hand side of (\ref{dQbh}) can be written as
\be \label{TdSbs}
\delta \int_\mathcal{B} \Q_\xi = \frac{\kappa}{2\pi} \delta S_{\mathrm{bh}} = T \, \delta S_{\mathrm{bh}}
\ee
As in manifestly covariant theories, the black hole temperature $T$ can be identified with $\kappa/2\pi$. The black hole entropy $S_{\mathrm{bh}}$ has two contributions,
\be
S_{\mathrm{bh}} = S_{\mathrm{cov}} + S_{\mathrm{CS}}
\ee 
where $S_{\mathrm{cov}}$ comes from the manifestly covariant part of the Lagrangian 
$\LL_{\mathrm{cov}}$, and $S_{\mathrm{CS}}$ from the noncovariant part of the Lagrangian 
$\LL_{\mathrm{CS}}$. $S_{\mathrm{cov}}$ is obtained by putting $\mathcal{L}_{\mathrm{cov}}$ in 
Wald's formula (\ref{waldent}), while $S_{\mathrm{CS}}$ can be obtained by using 
$\Q_\xi^{(1)}$ calculated in Section \ref{sec:covphsp}. 
For example, from (\ref{Qxi1}) follows that the contribution from the pure gravitational Chern-Simons term 
(\ref{LpCS}) to the black hole entropy formula is
\be \label{SCSpt}
S_{\mathrm{CS}} = 2\pi n(n-1) \int_0^1 dt \int_{\mathcal{B}} P_n(\epsilon,\Gam,\R_t^{n-2})
\ee

It is interesting that C-S contribution to the entropy formula (\ref{SCSpt}), 
and analogous relations for mixed cases, 
can also be put in the form analogous to the Wald formula (\ref{waldent}). 
To see this let us write Wald entropy formula (\ref{waldent}) in the equivalent way\footnote{This follows from
\be
\epsilon_D \big|_\mathcal{B} = \epsilon \wedge \tilde{\epsilon}_{D-2}\0
\ee
where $\epsilon_D$ is the volume $D$-form, $\tilde{\epsilon}_{D-2}$ is the induced volume 
$(D-2)$-form on $\mathcal{B}$, and $\epsilon$ is 2-form binormal to $\mathcal{B}$.}
\be \label{swDform}
S_W = 2\pi \int_{\mathcal{B}} \epsilon^a_{\;\; b}
 \left( \frac{\delta \LL_{\mathrm{cov}}}{\delta R^a_{\;\; b\mu\nu}} \right)_{\mu\nu\rho_1\cdots\rho_{D-2}}
\ee
For C-S contribution to the entropy we obtain
\be \label{genwaldent}
S_{\mathrm{CS}}
 = 2\pi \int_{\mathcal{B}} \epsilon^a_{\;\; b} \frac{\partial \LL_{\mathrm{CS}}^{(n)}}{\partial \R_t{}^a{}_b}
 = 2\pi \int_{\mathcal{B}} \epsilon^a_{\;\; b}
 \left( \frac{\delta \LL_{\mathrm{CS}}}{\delta R_t{}^a{}_{b\mu\nu}} \right)_{\mu\nu\rho_1\cdots\rho_{D-2}}
\ee
and we see that the only difference with (\ref{swDform}) is that one has to consider the variation with respect to
$\R_t$ instead of $\R$. 
Here it is understood that variation with respect to $\R_t$ acts inside the $t$ integral present in Lagrangian.
The Lagrangians used here are of the form $\LL_{\mathrm{gCS}}^{(i)}$ from (\ref{LCSgen}) which means that there will 
always be exactly one 
$\CS$ factor, and consequently exactly one $t$ integral.

As we show in Appendix \ref{app:Kruskal} (see e.g.\ relation (\ref{gam2bs})) using Kruskal-type coordinates enables us to substitute $\R_t$ with $t\R$ in (\ref{SCSpt}). We then obtain
\be \label{SCSp}
S_{\mathrm{CS}} = 2\pi n \int_{\mathcal{B}} P_n(\epsilon,\Gam,\R^{n-2})
\ee
In the mixed case 1 (\ref{mixCS}) one gets
\be
S_{1,\mathrm{mixCS}} &=& 2\pi m \int_{\mathcal{B}} P_m(\epsilon,\Gam,\R^{m-2}) P_k(\mathbf{A}) \label{mCSp1}\\
S_{2,\mathrm{mixCS}} &=& 2\pi m \int_{\mathcal{B}}  P_m(\epsilon,\R^{m-1}) \CS_{\mathrm{CS}}^{(k)}(\A)\label{mCSp2}
\ee
In the above integrals it is of course understood that $\Gam,\R$ and $\A$ are replaced by their pull-backs 
by the immersion of ${\mathcal{B}}$ in space-time. The pull-backs of $\Gam$ and $\R$ have in general components
in all of the Lie algebra of $SO(1,D-1)$. However the use of Kruskal-type coordinates, adapted to the black hole geometry,
entails that the only 
contribution to such terms comes when the tensor indices are in the 2-dimensional normal bundle of the $(D-2)$-dimensional bifurcation surface $\mathcal{B}$. In other words 
\be \label{ourTach}
\str(\epsilon\,\Gam\,\R^{2k}) = 2\, \Gam_N \, \R_N^{2k} \;,
\ee
where
\be \label{GamN}
\Gam_N \equiv \frac{1}{2} \tr(\epsilon \Gam) \;, \qquad
\R_N \equiv \frac{1}{2} \tr(\epsilon \R)
\ee
Later on we shall use (\ref{ourTach}) to analyse the issue of coordinate independence of black hole entropy in 
theories with CS terms in the action\footnote{Formula (\ref{mCSp2}) is of the covariant type entropy formulas, since $\A$ is simply a one-form from the diff point of view. Therefore from now on, when talking about covariantization, we will refer only to formulas (\ref{SCSp},\ref{mCSp1}).}.

To derive (\ref{ourTach}) 
we note from (\ref{lambdaKrus}) and (\ref{lambf}) that the components of the binormal in Kruskal coordinates are
\be 
\epsilon{}^U{}_U = 1 \;, \qquad \epsilon{}^V{}_V = -1
\ee
then by using (\ref{nabxicov})-(\ref{Riebscov}), it follows 
\be \label{ourexp1}
\str(\epsilon\,\Gam\,\R^{n-2})
 &=& \Gam{}^U{}_U (\R{}^U{}_U )^{n-2} 
  - \Gam{}^V{}_V (\R{}^V{}_V )^{n-2}  \0 
\ee
On the bifurcation surface we have $\R{}^U{}_U = -\R{}^V{}_V = \R_N$. Since $n$ is even it follows
\be \label{ourexp}
\str(\epsilon\,\Gam\,\R^{n-2})
 &=& (\Gam{}^U{}_U - \Gam{}^V{}_V ) (\R_N )^{n-2}  \0 \\
 &=& 2 \, \Gam_N \, \R_N^{n-2}
\ee
where $\Gam$ and $\R$ denote the pull-backs on ${\mathcal{B}}$. The `miracle' of Kruskal-type coordinates
is that (locally) they kill all the components of $\Gamma$, except those in the normal bundle directions.

Using (\ref{ourexp}) the formula for the entropy becomes
\be \label{SCSp2}
S_{\mathrm{CS}} = 4\pi n \int_{\mathcal{B}} \Gam_N \, \R_N^{n-2}
\ee

\subsection{Covariance of black hole entropy}
\label{ssec:covariance}

In Section \ref{ssec:entKrus} we have obtained a formula for the black hole entropy by using a special
type of coordinate system --  Kruskal-type coordinates. 
This raises immediately the question of the validity of the same formula in a generic coordinate system. Beside, in practical calculations one rarely uses
Kruskal-type coordinates, and ``Schwarzschild-type" coordinates (which, on the other hand, are not even regular at $\mathcal{B}$) are typically the most efficient. 
In reality it is possible to derive the same formula (\ref{SCSp2}) in a purely geometric and coordinate independent way: see Appendix \ref{app:spinconn}. 
The geometry we have to do with in this section is determined by the presence of a 
submanifold ${\mathcal{B}}$, which naturally breaks the group of diffeomorphisms into the direct product of $Diff_{\mathcal{B}}$ and
the diffeomorphisms that leave ${\mathcal{B}}$ pointwise unchanged. The latter can be interpreted as gauge transformations
of the ${\mathcal{B}}$ normal bundle, $N{\mathcal{B}}$. For our problem the relevant invariance to consider is the invariance of the entropy formula under this product group\footnote{One can consider also more general diffeomorphisms, see for instance \cite{BR}, but we will ignore such kinds of subtleties here.}.

Although this problem can be dealt with in more generality (see Appendix \ref{app:spinconn}), it is interesting to examine the question of the black hole entropy formula(s)' invariance 
from the point we left it in the previous subsection, i.e. the more popular point of view of
Kruskal coordinates. Our purpose in this subsection is to 'covariantize' (\ref{SCSp2}), that is to find a covariant formula for entropy that coincides with (\ref{SCSp2}) in Kruskal coordinates. This covariant formula will turn out to coincide with that found in  Appendix \ref{app:spinconn}.

In the previous subsection we have obtained that the contribution to the black hole entropy due to the
presence of CS terms in the action, calculated in Kruskal coordinates, will be in the most general case a sum of terms 
of the form
\be \label{CSentcontr1}
2\pi (2k_1 + 2) \int_{\mathcal{B}} \str(\epsilon\, \Gam\, \R^{2k_1}) \str(\R^{2k_2}) \cdots \str(\F^{m_1}) \cdots
 \ ,\0\\
4 \sum_i k_i + 2 \sum_j m_j = D - 3
\ee
We have seen that this can be reduced to\footnote{Once again we recall that by $\Gam$ and
$\R$ in (\ref{CSentcontr1}) we mean the pull-back to ${\mathcal{B}}$ of the corresponding forms in space time.}
\be \label{CSentcontr2}
2\pi (2k_1 + 2) \int_{\mathcal{B}} \Gam_N \, \R_N^{2k_1} \str(\R^{2k_2}) \cdots \str(\F^{m_1}) \cdots
\ee

Now, in Kruskal coordinates we have
\be \label{eGamNom}
q^\nu_\mu \, \Gam_{N\nu} \bifeq q^\nu_\mu \, u^\rho \nabla_\nu v_\rho
 \equiv {\mathbf{\omega}}_\mu
\ee
where $v^\mu$ and $u^\mu$ are null vector fields (normalized to $u_\mu v^\mu = -1$) normal to $\mathcal{B}$ defined in (\ref{uvdef}) in Appendix \ref{app:Kruskal}, and $q^\mu_\nu$ is a tangential projector on $\mathcal{B}$ (first fundamental tensor of $\mathcal{B}$) defined in Appendix \ref{ssec:surfaces}. The second equality is simply the definition of $\boldsymbol{\omega}$, which is nothing but the connection of the normal bundle, see (\ref{omega1}).
 
Next we express also $\R_N$ in terms of $\boldsymbol{\omega}$.
Using (\ref{OmRN}) and (\ref{Om2om1}) we get
\be \label{RNcov}
q_\mu^\rho \, q_\nu^\sigma \, (\R_N)_{\rho\sigma} \bifeq {\mathbf{\Omega}}_{\mu\nu}
 = q_\mu^\rho \, q_\nu^\sigma (d{\mathbf{\omega}})_{\rho\sigma}
\ee
where $\boldsymbol{\Omega}$ is the normal curvature 2-form. 

What is important here is that: (a) by definition, $\boldsymbol{\omega}$ is a 1-form, 
$\boldsymbol{\Omega}$ is a 2-form on $\mathcal{B}$ (b) inside the integral 
(\ref{CSentcontr2}) one can substitute $\Gam_{N\mu}$ with $q^\nu_\mu \, \Gam_{N\nu}$, and also
$(\R_N)_{\mu\nu}$ with $q_\mu^\rho q_\nu^\sigma (\R_N)_{\rho\sigma}$, because integration over 
$\mathcal{B}$ implies tangential projection by $q_\mu^\nu$. In particular we remark that
$\boldsymbol{\omega}$ behaves as a 1-form under 
the diffeomorphisms of $\mathcal{B}$ and $d$ is the exterior derivative on $\mathcal{B}$.
So the integral (\ref{CSentcontr2}), with $\Gam$ replaced by $\mathbf{\omega}$,
is well defined and covariant with respect to the diffeomorphisms
of $\mathcal{B}$. In summary,
we should replace (\ref{CSentcontr2}) with 
\be \label{CSentcov}
2\pi (2k_1 + 2) \int_{\mathcal{B}} \boldsymbol{\omega} \, \boldsymbol{\Omega}^{2k_1} \,
 \str(\R^{2k_2}) \cdots \str(\F^{m_1}) \cdots
\ee
and use this form of the integral to prove its invariance under normal bundle gauge transformations. This is simple.
The structure group of the normal bundle is SO(1,1). The corresponding gauge transformations are given
by rescalings $v^\mu \to f v^\mu$, $u^\mu \to u^\mu/f$ (see Appendix~\ref{ssec:surfaces}), where $f(x)$ is an arbitrary positive function. 
From (\ref{eGamNom}) it follows that $\boldsymbol{\omega}$ transforms as
\be \label{tomgauge}
\boldsymbol{\omega} \rightarrow \boldsymbol{\omega} + d \ln f
\ee
From this the transformation rule
\be \label{CSnbtrans}
\boldsymbol{{\omega}} (d\boldsymbol{{\omega}})^{2k} \to 
\boldsymbol{{\omega}} (d\boldsymbol{{\omega}})^{2k} + d(\ldots)
\ee
follows. Finally, we see that, as a consequence of (\ref{dLCS}), the formula for entropy (\ref{CSentcov}) is invariant under the gauge transformation (\ref{tomgauge}).

In conclusion, we have shown that the expressions for the CS black hole entropy terms (\ref{CSentcov}) are invariant.

\subsection{BH entropy in Schwarzschild-type coordinates}
\label{ssec:BHSchwarz}

In the previous subsection it was shown how the contribution to the black hole entropy formula from the CS Lagrangian terms can be written in the manifestly diff-invariant form. Now we turn to the question of practical calculations. In practice, calculations are typically performed in coordinate systems in which symmetries are maximally exploited, which typically means that a maximal set of linearly independent Killing vector fields are used as generators for coordinates. We call such coordinate systems ``Schwarzschild-type" (because the coordinates used in the Schwarzschild solution are one such example). The bonus of these coordinate systems is that if there are $N$ Killing vectors, then the components of the metric depend only on $(D-N)$ coordinates, which substantially simplifies the calculations (in particular, to solve the equations of motion). In our calculations, the obvious simplification comes from the property that in Schwarzschild-type coordinates, by construction, the components of all Killing vectors are constant everywhere, and this is true in particular for the horizon generating Killing vector $\xi^a$. This means that 
$\Lambda^a{}_b \equiv \partial_b \xi^a =0$, and so also $d\Lambda=0$, everywhere. This implies 
$\Sigma_\xi = 0$, like in Kruskal-type coordinates.

However, there is a price to pay. Schwarzschild-type coordinates are by construction \emph{singular} on the bifurcation surface $\mathcal{B}$ (in fact, on the whole horizon $\mathcal{H}$). The constancy of the components $\xi^a$, due to (\ref{xibf0}), implies that the metric tensor components, calculated in Schwarzschild-type coordinates, must be singular on $\mathcal{B}$. This means, for example, that 
(\ref{lambf}) and (\ref{ixiTHB}) are not valid in these coordinates. Therefore one has to be careful when using Schwarzschild-type coordinates to compute quantities at the horizon, and the entropy in Wald formalism is one such quantity. In manifestly covariant theories this singular behavior does not pose a problem - physical quantities typically can be put in diff-invariant form which allows one to perform calculations slightly outside the horizon and then take the horizon limit which is well-defined for diff-scalars. When CS terms are present, manifest diff-covariance is broken and, as a consequence, one has to make sure how and when one can calculate with Schwarzschild-type coordinates. One of the benefits of the covariantization from the previous subsection is that we can safely use Schwarzschild-type coordinates in calculating the entropy terms in the form (\ref{CSentcov}).

What we want to show here is that in Schwarzschild-type coordinates the CS contributions (\ref{CSentcov}) to the entropy can be calculated directly from
\be \label{CSentSch}
\int_{\mathcal{B}} \Gam_N \R_N^{2k} \, \tr(\R^{2k_2}) \cdots \tr(\F^{m_1}) \cdots
\ee
We shall now prove this explicitly for the case of stationary black holes. Schwarzschild-type coordinates, and their connection to Kruskal-type coordinates, are presented in Appendix \ref{app:Sch}. It was shown in \cite{Racz:1992bp} that ``natural" (fixed-time) Cauchy surfaces $\mathcal{C}_t$ (defined by $t=\mbox{const}$) are ending on the  bifurcation surface $\mathcal{B}$, which is convenient for analysis of the first law of thermodynamics, which follows from (\ref{dQKill})-(\ref{dQbh}). As we show in detail in Appendix \ref{app:Sch} 
\be \label{GNSK}
\left( \Gam_N \right)_{\textrm{Schwarzschild}} \bifeq \left( \Gam_N \right)_{\textrm{Kruskal}}
\ee
where in subscript we denote the coordinate system in which the calculation is performed. As all other factors in (\ref{CSentSch}) are manifestly covariant, from (\ref{GNSK}) it follows that it does not matter whether we calculate (\ref{CSentSch}) in Schwarzschild-type or Kruskal-type coordinates. This completes the proof.

\section{Conclusion}

In this paper we have shown that the addition of CS terms to a gravitational action can be dealt in a clear, if
not straightforward, way. We have extended the covariant phase space formalism to such noncovariant situations and found
a general formula for the black hole entropy. This formula looks very similar to the covariant Wald formula, but with a significant difference that seems to render it non-covariant. By exploiting the geometry of the bifurcation surface we have however been able, by means of Kruskal coordinates, to reduce the entropy formula to a form that can be 'covariantized'. In other words we have found a covariant formula that reduces, in Kruskal coordinates, to the latter. The final 'covariantized' formula is covariant with respect to the diffeomorphisms of the bifurcation surface and to the gauge transformations of the normal bundle. We have outlined a derivation of this formula from the initial seemingly non-covariant form in a coordinate independent way, via the geometry of the relevant orthogonal bundles.

\acknowledgments 

One of us (L.B.) would like to thank the Yukawa Institute for Theoretical Physics, Kyoto, for the hospitality and financial
support during this research, and the Theoretical Physics Department, Univ.\ of Zagreb, for hospitality and financial support
during his visits there. I.S.\ would like to acknowledge the financial support of  CEI Fellowship Programme CERES.
Also, M.C., P.D.P., S.P.\ and I.S.\ would like to thank SISSA for hospitality and financial support during visits there and would also like to acknowledge support by the Croatian Ministry of Science, Education and Sport under the contract no.~119-0982930-1016.

\vspace{0.5cm}

\section*{Appendix}
\appendix

\vspace{0.5cm}

\section{Mathematical preliminaries}

\vspace{0.5cm}

We denote $\mathfrak{g}$-valued ($\mathfrak {g}$ being a Lie algebra)  $p$-forms as follows:
\be
\mathbf{A} = \mathbf{A}^a \, T^a = \frac{1}{p!} \, A^a_{\mu_1 \cdots \mu_p} \, T^a \, dx^{\mu_1} \wedge \dots \wedge dx^{\mu_p} \ \in \ \mathfrak{g} \otimes \Lambda^p(\mathscr{M})
\ee
Commutators between such objects are defined in the following manner
\be
[\mathbf{A}, \mathbf{B}] = \mathbf{A}^a \mathbf{B}^b \, [T^a, T^b] = \mathbf{AB} - (-1)^{pq} \, \mathbf{BA}
\ee
where $\mathbf{A}^a$ is $p$-form and $\mathbf{B}^a$ is $q$-form. In the case of the Christoffel 1-form this means that
\be
\tensor{[\Gam, \Gam]}{^a_b} = 2 \tensor{(\Gam^2)}{^a_b} = 2 \Gamma^a_{c\mu} \, \Gamma^c_{b\nu} \, dx^\mu \wedge dx^\nu
\ee
Throughout the paper we are using the standard notation,
\be
\R = d\Gam + \Gam^2 \qqd \Gam_t = t\Gam \qqd \R_t = d\Gam_t + \Gam_t^2 = t\R + (t^2 - t)\Gam^2
\ee
Covariant exterior derivatives are
\be
\Db = d + [\Gam , \, . \ ] \qqd \Db_t = d + [\Gam_t , \, . \ ]
\ee
and Bianchi identity reads $\Db \R =0$ and $\Db_t \R_t =0$.

We often use the identity
$$\dd_t \R_t = \R + (2t - 1) \Gam^2 = d\Gam + 2t \Gam^2 = d\Gam + [\Gam_t, \Gam] = \Db_t \Gam$$
and the variation of the Riemann 2-form,
$$\delta \R_t = \delta d \Gam_t + \delta \Gam_t \Gam_t + \Gam_t \delta \Gam_t = d\delta \Gam_t + [\Gam_t, \delta \Gam_t] = \Db_t \delta \Gam_t$$
The noncovariant part of the diff-variation $\delta_\xi$ is defined by
$$\hat{\delta}_\xi \Gam = d \Lambda$$
and implies
$$\hat{\delta}_\xi \Gam_t = t d\Lambda, \quad\quad\hat{\delta}_\xi \R = 0 \qqd \hat{\delta}_\xi \R_t = (t-1) [\Gam_t, d\Lambda] = (t-1) \Db_t d\Lambda$$

\vspace{1.0cm}

\noindent
Let $M_k(\cc)$ be a set of complex $k \times k$ matrices, and $P_n : \otimes^n \, M_k(\cc) \rightarrow \cc$ $n$-linear functions (linear in each of its arguments), which are\\

\noindent
$a)$ \ symmetric,
\be
P_n (A_1, \dots, A_i, \dots, A_j, \dots, A_n) = P_n (A_1, \dots, A_j, \dots, A_i, \dots, A_n)
\ee
for all $A_r \in M_k(\cc)$, $1 \le r \le n$, and\\

\noindent
$b)$ \ adjoint invariant with respect to the gauge group $G$,
\be
P_n (g^{-1} A_1 g, \dots, g^{-1} A_n g) = P_n (A_1, \dots, A_n)
\ee
for all so $g \in G$ and $A_r \in \mathfrak{g} = \textrm{Lie} \, G$.\\

\noindent
A map $P_n$ that satisfies these properties is called \emph{symmetric invariant polynomial}. An example of such polynomial is \emph{symmetrized trace},
$$\str \, (A_1, \dots, A_n) \, \equiv \, \frac{1}{n!} \, \sum_{\pi} \, \textrm{tr}\left(A_{\pi(1)} \cdots A_{\pi(n)}\right)$$
where $\pi$ denotes the permutations of the indices $\{1, \dots, n\}$. 

When some of the arguments of the invariant polynomial happen to be equal, we often use the following abbreviations,
$$P_n (\mathbf{A}, \mathbf{B}^{n-1}) \equiv P_n (\mathbf{A}, \mathbf{B}, \dots, \mathbf{B})$$
$$P_n (\mathbf{A}, \mathbf{B}, \mathbf{C}^{n-2}) \equiv P_n (\mathbf{A}, \mathbf{B}, \mathbf{C}, \dots, \mathbf{C})$$
$$\textrm{etc.}$$
It is easy to extend the domain of invariant polynomials from $\mathfrak{g}$ to $\mathfrak{g}$-valued $p$-forms on manifold $\mathscr{M}$, by
$$P_n (A_1 \eta_1, \dots, A_n \eta_n) \ \equiv \ P_n (A_1, \dots, A_n) \, \eta_1 \wedge \dots \wedge \eta_n \qqd \eta_i \in \Lambda^{p_i}(M)$$
We often use the following identity (proof can be found e.g.~in \cite{Nakahara}),
\be\label{dPn}
dP_n (A_1 \eta_1, \dots, A_n \eta_n) = \sum_{i=1}^n (-1)^{p_1 + \dots + p_{i-1}} \, P_n (A_1 \eta_1, \dots, D (A_i \eta_i), \dots, A_n)
\ee
where covariant derivative $D$ is defined as
$$D (A_i \eta_i) = A_i d\eta_i + [\theta, A_i \eta_i] = A_i d\eta_i + \theta^a_\mu A^b_i \, [T^a, T^b] \, dx^\mu \wedge \eta_i$$
for some $\mathfrak{g}$-valued 1-form $\theta$.\\

\section{Some exact formulae}
\label{app:exact}

In this Appendix we derive some exact relations, which are used in the text or will be useful
later on. Here exact means that they are true
also {\it off shell}. The purpose of this appendix is also to show some of the techniques
we use in deriving our formulas.

\subsection{Noether current}
\label{app:puregrav}

Let us start from 
\be
\J_{\xi} = \Th^{cov}_{\xi}+\Th^{nc}_{\xi} -i_{\xi}\LL_{\mathrm{pCS}}-\X_{\xi}\label{Jxi}
\ee 
where
\be
\Th^{nc}_{\xi}\equiv \Th^{nc}(\delta\Gam= (i_{\xi}d+di_{\xi})\Gam + D\Lambda)\0
\ee
Using the methods from section \ref{sec:eom} we have 
\be
\Th^{nc}(\delta\Gam= D\Lambda)-\X_{\xi}= d\mathbf{q}_{1,\xi}-n P_n(\Lambda,\R^{n-1})\label{dQ1}
\ee
where
\be
\mathbf{q}_{1,\xi}(\Gam)= n(n-1)\int_0^1dt P_n(\Lambda,\Gam,\R_t^{n-2})\label{Q1}
\ee
Now let us use
\be
i_{\xi}\R_t = L_{\xi}\Gam_t -D_t(i_{\xi} \Gam_t)\label{ixiRt}
\ee
where $L_{\xi}=i_{\xi}d+di_{\xi}$. We have
\be
i_{\xi}\LL_{\mathrm{pCS}}&=& n\int_0^1dt P_n(i_{\xi}\Gam, \R_t^{n-1})-n(n-1) \int_0^1dt P_n(\Gam, L_{\xi}\Gam_t -D_t(i_{\xi}\Gam_t), \R_t^{n-2})\0\\
&=&n\int_0^1dt P_n(i_{\xi}\Gam, \R_t^{n-1})-n(n-1) \int_0^1dt P_n(\Gam, L_{\xi}\Gam_t, \R_t^{n-2})\0\\
&&- n(n-1) d\int_0^1dt P_n(i_{\xi}\Gam,\Gam_t, \R_t^{n-2})+n\int_0^1dt\, t \frac d{dt}P_n(i_{\xi}\Gam, \R_t^{n-1})\0\\
&=&nP_n(i_{\xi}\Gam ,\R^{n-1})+n(n-1)\int_0^1dt P_n(L_{\xi}\Gam,\Gam_t,\R_t^{n-2})\0\\
&&-n(n-1) d\int_0^1dt P_n(i_{\xi}\Gam,\Gam_t,\R_t^{n-2})\label{LCSdQ2}
\ee
On the other hand
\be
\Th^{nc}(\delta\Gam= (i_{\xi}d+di_{\xi})\Gam )=n(n-1)\int_0^1dt P_n(L_{\xi}\Gam,\Gam_t,\R_t^{n-2})\0
\ee
It follows that
\be
\Th^{nc}(\delta\Gam= (i_{\xi}d+di_{\xi})\Gam )-i_{\xi}\LL_{\mathrm{pCS}}=-
nP_n(i_{\xi}\Gam,\R^{n-1})+ d \mathbf{q}_{2,\xi}\label{ThetadQ2}
\ee
where
\be
\mathbf{q}_{2,\xi}=n(n-1)\int_0^1dt P_n(i_{\xi}\Gam,\Gam_t,\R_t^{n-2}) \label{Q2}
\ee
So, defining,
\be
\mathbf{q}_{\xi}(\Gam)&=& \mathbf{q}_{1,\xi}(\Gam)+\mathbf{q}_{2,\xi}(\Gam)\0\\
&=& 
n(n-1)\left[\int_0^1dt P_n(\Lambda,\Gam,\R_t^{n-2})+\int_0^1dt P_n(i_{\xi}\Gam,\Gam_t,\R_t^{n-2})\right] \label{qxi}
\ee
we have
\be
\J_{\xi} = \Th^{cov}_{\xi}- nP_n(i_{\xi}\Gam +\Lambda,\R^{n-1})+ d \mathbf{q}_{\xi}\label{Jdq}
\ee

The obstruction to the off-shell exactness of $\J_{\xi}$ is $\Th^{cov}_{\xi}- nP_n(i_{\xi}\Gam +\Lambda,\R^{n-1})$. This obstruction is removed by putting the system on-shell (see below). This
will allow us to define $\Q_\xi$.

One can derive similar exact equations for the mixed cases.

\subsection{Derivation of $\mathbf{\Sigma}_\xi$}
\label{app:Sigma}

We give here the derivation of $\mathbf{\Sigma}_\xi$ in the purely gravitational case.
\be
&&\hat \delta_\xi \Th - \delta \X_\xi \label{B2}\\
&=& - \hat \delta_\xi \left( n(n-1)\int_0^1 dt P_n(\Gam, \delta \Gam_t,
\R_t^{n-2})\right)- \delta\left(  n(n-1) \int_0^1 dt P_n(d\Lambda,\Gam,\R_t^{n-2})\right)\0\\
&=& - n(n-1)\Bigl{(}\int_0^1 dt\, P_n(d\Lambda,\delta \Gam_t,\R_t^{n-2})+(n-2)(t^2-t)P_n(\Gam,\delta \Gam_t,[d\Lambda,\Gam],\R_t^{n-3})\Big{)}\0\\
&&- n(n-1)\int_0^1 dt\,(t-1) \Big{(} P_n(d\Lambda,\delta \Gam, \R_t^{n-2})+(n-2) P_n(d\Lambda,\Gam, D_t\delta \Gam_t,\R_t^{n-3})\Big{)}\0\\
&=& -n(n-1)(n-2) d\int_0^1 dt\,(t-1) P_n(d\Lambda,\Gam,\delta \Gam_t,\R_t^{n-3})\0\\
&&-n(n-1)(n-2) \int_0^1 dt\,(t-1) P_n(d\Lambda, \frac {d \R_t}{dt},\delta\Gam_t,\R_t^{n-3})\0\\
&& -n(n-1)\int_0^1 dt\,(2t-1) P_n(d\Lambda,\delta \Gam, \R_t^{n-2})
- n(n-1)\int_0^1 dt\,P_n(d\Lambda, \delta \Gam_t, \R_t^{n-2})\0\\
&=& -n(n-1)(n-2) d\int_0^1 dt\,(t-1) P_n(d\Lambda,\Gam,\delta \Gam_t,\R_t^{n-3})\0\\
&&-n(n-1)(n-2) \int_0^1 dt\,\frac d{dt} \Bigl{(}(t-1) P_n(d\Lambda,\delta \Gam_t,\R_t^{n-3})\Big{)}\0\\
&=& -n(n-1)(n-2) d\int_0^1 dt\,(t-1) P_n(d\Lambda,\Gam,\delta \Gam_t,\R_t^{n-3})\0
\ee
That is 
\be
\Sig_{\xi} = -n(n-1)(n-2) \int_0^1 dt\,(t-1) P_n(d\Lambda,\Gam,\delta \Gam_t,\R_t^{n-3})\label{Sigma}
\ee

The generalization to the mixed cases is straightforward.

\section{Relations between Noether current and charge}
\label{app:Qcalc}

In this appendix we present various ways of deriving the relation 
\be\label{JdQxi}
\J_\xi\approx d\Q_\xi
\ee
from the on-shell closedness of $\J_\xi$, $d\J_\xi\approx 0$. In (\ref{JdQxi}) both
$J_\xi$ and $Q_\xi$ are local expressions of the fields and their derivatives, and what is remarkable of the relation is of course the specification that it holds on shell. Without this specification the relation would simply be the statement of the local Poincar\'e lemma.
Modding out the equations of motion requires a more sophisticated version of the local 
Poincar\'e lemma. This problem has been tackled and solved by Wald, \cite{Wald1}, in a constructive way. But the same problem had been dealt with and solved in a more mathematical language, using the cohomology of the so-called variational bi-complex. In fact solving this
problem is equivalent to finding the homotopy operator(s) showing the exactness of such  
bi-complex. On the other hand Wald's constructive approach is tantamount to constructing the relevant homotopy operator(s).

In this Appendix we follow a line in between. We use a constructive approach, by introducing
a simple practical method to invert eq.(\ref{JdQxi}). This method relies on a new bi-complex 
that can be embedded in the standard variational bi-complex, thus transforming it into an  
variational tri-complex. Its exactness completely justifies our heuristic approach.

\subsection{Bicomplex formulation of Wald's Lemma}
\label{app:bicomplex}

In the cases of our interest, $\J_\xi$ is always of the form
\be \label{Jour}
\J_\xi = d\partial_a \xi^b \, \mathbf{X}{}^a{}_b + \partial_a \xi^b \, \mathbf{A}^{(1)}{}^a{}_b
 + \xi{}^b \mathbf{A}^{(0)}{}_b
\ee
where $\mathbf{X}$, $\mathbf{A}^{(1)}$ and $\mathbf{A}^{(0)}$ are tensor-valued forms (which are local functionals of dynamical fields $\phi$). We know that on-shell (i.e., when $\phi$ satisfy EOM) $\J_\xi$ is closed, so
\be \label{JdQ}
d\J_\xi = 0 \qquad \Longrightarrow \qquad \J_\xi = d\Q_\xi
\ee
Our goal is to calculate $\Q_\xi$.

We now show that Wald's constructive result \cite{Wald1} can be cast into an elegant cohomological formulation. In this subsection we will simply ignore the specification that
the relations we are dealing with are true on shell. This will be justified in the next
subsection, by inserting the following construction into an exact variational bicomplex.

Following Wald, we notice that in all formulae indices which are contracted with partial derivatives of $\xi$ appear only in symmetrized combinations, so we work only with tensor-valued $p$-forms of the form
\be
H{}_{\mu_1\cdots\mu_p}{}^{a_1 \cdots a_j}{}_b =
 H{}_{[\mu_1\cdots\mu_p]}{}^{(a_1 \cdots a_j)}{}_b
\ee
which we denote in short $H^j_p$, and in general call ${}^{j}_p$-tensors. By definition, when $j<0$ or
$p<0$ then $H^j_p = 0$. The exterior differential $d$ changes ${}^{j}_p$-tensor to ${}^{j}_{p+1}$-tensors. We define two additional operations on ${}^{j}_p$-tensors. One is the "$g$-wedge" operation, $\gwedge{}$, defined by 
\be
\left( \gwedge{} \, H^j_p \right){}_{\mu_1\cdots\mu_{p+1}}{}^{a_1 \cdots a_{j+1}}{}_b \equiv
 (p+1) \, g^{(a_1}_{[\mu_1} H{}_{\mu_2\cdots\mu_{p+1}]}{}^{a_2 \cdots a_{j+1})}{}_b
\ee
($g^a_\mu = \delta^a_\mu$ is the metric tensor) 
which transform ${}^{j}_p$-tensors into ${}^{j+1}_{p+1}$-tensors. We also define a "$g$-contraction" operation, $i_g$, by
\be
\left( i_g H^j_p \right){}_{\mu_1\cdots\mu_{p-1}}{}^{a_1 \cdots a_{j-1}}{}_b \equiv
 j \, g^\nu_c (H^j_p){}_{\nu\mu_1\cdots\mu_{p-1}}{}^{c a_1 \cdots a_{j-1}}{}_b
\ee
which transform ${}^{j}_p$-tensors into ${}^{j-1}_{p-1}$-tensors. The anticommutator of $\gwedge{}$ and 
$i_g$ satisfies
\be\label{gg}
(i_g \gwedge{} + \gwedge{} i_g {}) \, \mathbf{H}^j_p  = (D - p + j) \, \mathbf{H}^j_p 
\ee
which means that $(D - p + j)^{-1} i_g{}$ is the homotopy operator corresponding to $\gwedge{}$. From (\ref{gg}) it follows that if 
$\mathbf{H}^j_p$ satisfies $\gwedge{}\mathbf{H}^j_p = 0$, then (with the exception of the case in which $j=0$ and $p=D$)
\be \label{ghol}
\mathbf{H}^j_p = \gwedge{}\left( (D - p + j)^{-1} i_g\mathbf{H}^j_p \right)
\ee

The operations $d$ and $\gwedge{}$ satisfy the following relations
\be
d^2 = 0 \;, \quad \gwedge{}^2= 0 \;, \quad \gwedge{} d + d \gwedge{} = 0
\ee
which means that they define a (commutative) bi-complex. In the following we will use
\be
d \mathbf{H} = 0 \quad \Longleftrightarrow \quad \mathbf{H} = d \mathbf{B} \label{dcoh} \\
\gwedge{} \mathbf{H} = 0 \quad \Longleftrightarrow \quad \mathbf{H} = \gwedge{} \mathbf{B} \label{gcoh}
\ee
The second relation has been motivated a few lines above. The first is the already mentioned local Poincar\'e lemma. Its on-shell validity relies on the construction of a suitable homotopy operator which will be discussed below.

Let us now apply this formalism to prove a generalization of (\ref{JdQxi}). Let us start from a $p$-form $\J_\xi$
\be \label{Jgenl}
(J_\xi)_{\mu_1\cdots\mu_p} = \sum_{j=0}^k \left( \partial_{a_1}\cdots\partial_{a_j} \xi^b \right)
 A{}^{(j)}_{\mu_1\cdots\mu_p}{}^{a_1 \cdots a_j}{}_b,
\ee
and introduce a more compact notation for (\ref{Jgenl})
\be \label{Jgen}
\J_\xi = \sum_{j=0}^k \left( \partial^j \xi \right) \mathbf{A}{}^{(j)}
\ee
which enables us to get rid of all tensor indices in the formalism. In formulas such as
(\ref{Jgen}) it is understood  that $\mathbf{A}^{(j)}$ are ${}^j_p$-tensors. We assume now  that $\J_\xi$ is closed, as in (\ref{JdQ}). This enforces the following descent equations
\be
d \mathbf{A}^{(0)} &=& 0 \0 \\
&\vdots& \0 \\
d \mathbf{A}^{(j)} &=& - \gwedge{} \mathbf{A}^{(j-1)} \;\;, \quad j = 1,\ldots,k  \label{dJchain} \\
&\vdots& \0 \\
0 &=& \gwedge{}\mathbf{A}^{(k)} \0
\ee
The last equality implies that $\mathbf{A}^{(k)}$ has to be $g$-exact, i.e., that there exists a 
${}^{k-1}_{p-1}$-tensor $\mathbf{X}$ such that
\be \label{AkgX}
\mathbf{A}^{(k)} = \gwedge{} \mathbf{X}
\ee

We want to find a $(p-1)$-form $\Q_\xi$ satisfying (\ref{JdQxi}). Using linearity and smoothness in $\xi$, we
can conclude that $\Q_\xi$ can be written as
\be \label{Qgen}
\Q_\xi = \sum_{j=0}^\infty \left( \partial^j \xi \right) \mathbf{P}{}^{(j)}
\ee
where $\mathbf{P}{}^{(j)}$ are ${}^j_{p-1}$-tensors that we want to determine by using (\ref{JdQ}). Inserting
(\ref{Qgen}) and (\ref{Jgen}) in (\ref{JdQxi}) we obtain
\be
d \mathbf{P}^{(0)} &=& \mathbf{A}^{(0)} \label{dQg0} \\
\gwedge{} \mathbf{P}^{(j-1)} + d \mathbf{P}^{(j)} &=& \mathbf{A}^{(j)} \;, \quad j = 1,\ldots,k-1 \label{dQg1} \\
\gwedge{}\mathbf{P}^{(k-1)} + d \mathbf{P}^{(k)} &=& \gwedge{}\mathbf{X} \label{dQg2} \\
\gwedge{} \mathbf{P}^{(j-1)} + d \mathbf{P}^{(j)} &=& 0 \;\;, \qquad j = k+1,\ldots  \label{dQg3}
\ee
In (\ref{dQg2}) we have used (\ref{AkgX}). Observe that from (\ref{dQg2}) it follows that 
$\mathbf{P}^{(k)}$ must have form
\be \label{P2form}
\mathbf{P}^{(k)} = d \mathbf{Z} + \gwedge{} \mathbf{Z}'
\ee
Now we use the fact that solution to the descent equations is not unique due to the freedom 
\be \label{Qred}
\Q_\xi \sim \Q_\xi + d \mathbf{R}_\xi 
\ee
where $\mathbf{R}_\xi$ is arbitrary $(p-2)$-form. We take
\be
\mathbf{R}_\xi = \sum_{j=0}^\infty (\partial^j \xi) \mathbf{R}^{(j)}
\ee
where $\mathbf{R}^{(j)}$ are ${}^j_{p-2}$-tensors we can freely choose. Therefore
\be
d \mathbf{R}_\xi = \sum_{j=0}^\infty (\partial^j \xi) \left( \gwedge{} \mathbf{R}^{(j-1)} + d \mathbf{R}^{(j)} \right),
\ee
which means that we can freely redefine $\Q_\xi$ so that
\be \label{Predef}
\mathbf{P}^{(j)} \sim \mathbf{P}^{(j)} + \gwedge{}\mathbf{R}^{(j-1)} + d \mathbf{R}^{(j)}
\ee
It then follows from (\ref{Predef}), (\ref{P2form}) and (\ref{dQg3}) that by properly fixing $\mathbf{R}^{(j)}$,
$j \ge 1$, we can make
\be\label{Pj=0}
\mathbf{P}^{(j)} = 0 \;\;, \quad j = k,k+1,\ldots
\ee
This means that in (\ref{Qgen}) one can terminate the sum with $j \le (k-1)$, as claimed by Wald.

To avoid unnecessary complications, we will explicitly solve the descent equations 
(\ref{dQg0})-(\ref{dQg3}) in the case we are primarily interested in, i.e., when $p = D-1$ and $k = 2$ in (\ref{Jgen}). Then the equations become
\be
d \mathbf{P}^{(0)} &=& \mathbf{A}^{(0)}  \label{dQch0} \\
\gwedge{}\mathbf{P}^{(0)} + d \mathbf{P}^{(1)} &=& \mathbf{A}^{(1)} \label{dQch1} \\
\gwedge{}\mathbf{P}^{(1)} &=& \gwedge{}\mathbf{X} \label{dQch2}
\ee
From (\ref{dQch2}) it directly follows that
\be \label{P1Xg}
\mathbf{P}^{(1)} = \mathbf{X} + \gwedge{} (\ldots)
\ee
But from (\ref{Predef}) it is easy to see that by appropriately choosing $\mathbf{R}^{(0)}$ we can discard $\gwedge{}$-exact terms in (\ref{P1Xg}), so we are left with
\be \label{P1sol}
\mathbf{P}^{(1)} = \mathbf{X}
\ee
Now we use (\ref{P1sol}) in (\ref{dQch1}) to obtain
\begin{displaymath}
\gwedge{}\mathbf{P}^{(0)} = \mathbf{A}^{(1)} - d \mathbf{X}
\end{displaymath}
which we further operate with $i_g$ from the left. Using (\ref{gg}) on the LHS we finally obtain
\be \label{P0sol}
\mathbf{P}^{(0)} = \frac{1}{2} \, i_g (\mathbf{A}^{(1)} - d \mathbf{X})
\ee
(\ref{P1sol})-(\ref{P0sol}) together constitute the searched for solution to (\ref{JdQxi}).  

Let us pause here to check consistency of this solution. This means that we must
obtain (\ref{dQch0})-(\ref{dQch2}) from (\ref{P1sol})-(\ref{P0sol}) using only the $d\mathbf{J}=0$ descent equations(\ref{dJchain}). Relation (\ref{dQch2}) is a trivial consequence of (\ref{P1sol}). Then, to obtain 
(\ref{dQch1}) we operate on (\ref{P0sol}) from the left, which gives
\be
\gwedge{} \mathbf{P}^{(0)} &=& \frac{1}{2} \, \gwedge{} i_g (\mathbf{A}^{(1)} - d \mathbf{X}) \0 \\
 &=& \mathbf{A}^{(1)} - d \mathbf{X} - i_g \gwedge{}(\mathbf{A}^{(1)} - d \mathbf{X}) \0 \\
 &=& \mathbf{A}^{(1)} - d \mathbf{X} - i_g (\gwedge{}\mathbf{A}^{(1)} + d \gwedge{}\mathbf{X}) \0 \\
 &=& \mathbf{A}^{(1)} - d \mathbf{X} - i_g (\gwedge{}\mathbf{A}^{(1)} + d \mathbf{A}^{(2)}) \0 \\
 &=& \mathbf{A}^{(1)} - d \mathbf{P}^{(1)} \label{gP0p}
\ee
and the last line is exactly (\ref{dQch1}). (In the second line we used (\ref{gg}), in the fourth line we have used the $j=2$ relation from (\ref{dJchain}), and in the fifth line (\ref{P1sol})). Finally, to obtain (\ref{dQch0}) we operate with the exterior derivative $d$ on Eq. (\ref{gP0p}) from the left, which gives
\be
d \gwedge{}\mathbf{P}^{(0)} = d \mathbf{A}^{(1)} = - \gwedge{}\mathbf{A}^{(0)}
\ee
In the last equality we have used the $j=1$ relation from (\ref{dJchain}). Using $dg = - gd$ and (\ref{gcoh}) 
we obtain
\be
d \mathbf{P}^{(0)} = \mathbf{A}^{(0)} + \gwedge{}(\ldots) = \mathbf{A}^{(0)}
\ee
which is exactly (\ref{dQch0}). (In last equality we used the fact that $(\ldots)$ would have to be 
${}^{-1}_{D-2}$-tensor, which does not exist (it is by definition identically zero).) This completes the consistency check.

\subsection{Contribution to $\Q_\xi$ from the pure gravitational C-S Lagrangian term}

We now apply the formal solution of the previous subsection to the contribution of pure gravitational CS  Lagrangian term to $\J_\xi$. We have
\be \label{Jexpl}
\J_{\xi} = \Th\cov_{\xi} + \Th\nc_{\xi} - i_{\xi} \LL_{\mathrm{pCS}} - \X_{\xi}  \0
\ee
where one uses (\ref{LpCS}), (\ref{Thetanc}), (\ref{Thetacov}), and (\ref{XiGamma}). Obviously $\J_\xi$ has the form (\ref{Jgenl}) with $k=2$ and $p = D-1 = 2n-2$, so from the previous subsection we know that we can take $\Q_\xi$ to have the following form
\be \label{QCS}
\Q_\xi = ( \partial \xi) \mathbf{P}{}^{(1)} + \xi \mathbf{P}{}^{(0)} \equiv \Q_\xi^{(1)} + \Q_\xi^{(0)}
\ee
where $\mathbf{P}{}^{(j)}$ are determined by (\ref{P1sol})-(\ref{P0sol}). Let us calculate them explicitly.

From $\Th\nc_{\xi}(\hat{\delta}\Gam) - \X_{\xi}$ we can extract $\mathbf{X}$, which when used in 
(\ref{P1sol}) and (\ref{QCS}) gives the part of $\Q_\xi$ proportional to 
$\Lambda{}^a{}_b = \partial_b\xi^a$ 
\be
\Q_\xi^{(1)} = n(n-1) \int_0^1 dt P_n(\Lambda,\Gam,\R_t^{n-2}) 
\ee

Though  the calculation of $\Q_\xi^{(0)}$ is in principle equally straightforward, it is technically much more involved due to the more complicated formulae (\ref{P0sol}) and also because there are several terms in (\ref{Jexpl}) which contribute to $\mathbf{A}^{(1)}$. The final result is
\be 
(Q_\xi^{(0)})_{\mu_1\cdots\mu_{2n-3}}
 &=& n(n-1) \int_0^1 dt P_n(i_\xi \Gam,\Gam_t,\R_t^{n-2})_{\mu_1\cdots\mu_{2n-3}} \0 \\
&& - n \, \xi_b \, K{}^{ab}{}_{a\mu_1\cdots\mu_{2n-3}}
 + \frac{n}{2}(2n-1) \xi_{[a} K{}^{ab}_{b\mu_1\cdots\mu_{2n-3}]}
\ee

\subsection{Contribution to $\Q_\xi$ of pure gravitational C-S term (alternative derivation)}

Since we know that the term linear in $\xi$ (that is $\Q_\xi^{(0)}$) of $\Q_\xi$ is irrelevant
for the purpose of calculating the entropy, we can follow a simplified path.
Starting from 
\be \0
\J_{\xi} = \Th\cov_{\xi} + \Th\nc_{\xi} - i_{\xi}\LL_{\mathrm{pCS}}   - \X_{\xi} 
\ee
we have already proved in Appendix B, eq.(\ref{Jdq},\ref{qxi}), that $\J_{\xi}$ can be written
\be
\J_\xi = \Th\cov_\xi - nP_n(\imath_\xi\Gam + \Lambda, \R^{n-1})+ d{\mathbf{q}}_\xi \label{JnotdQ}
\ee
where
\be
\mathbf{q}_{\xi}(\Gam) &=& \mathbf{q}_{1,\xi}(\Gam) + \mathbf{q}_{2,\xi}(\Gam)\0\\
&=& 
n(n-1)\left[\int_0^1 dt P_n(\Lambda, \Gam, \R_t^{n-2}) + \int_0^1 dt P_n(\imath_\xi\Gam, \Gam_t, \R_t^{n-2})\right] \label{Qxiapp}
\ee
Let us concentrate on $\Th\cov_\xi - nP_n(\imath_\xi\Gam + \Lambda, \R^{n-1})$. We know that this form must be closed on shell, and it is linear in $\xi$ and
$\partial \xi$ only. Therefore we can write it in the form
\be\label{thcovPn}
\Th\cov_\xi - nP_n(\imath_\xi\Gam + \Lambda, \R^{n-1})= \xi \A^{(0)}+ \partial \xi \A^{(1)}
\ee
It follows that
\be\label{descent}
&& d \A^{(0)}\approx 0\0\\
&& \gwedge{} \A^{(0)}+d \A^{(1)}\approx 0\0\\
&& \gwedge{} \A^{(1)}\approx 0\0
\ee
We expect of course 
\be\label{xiP0}
\Th\cov_\xi - nP_n(\imath_\xi\Gam + \Lambda, \R^{n-1})\approx d(\xi \mathbf{P}^{(0)})
\ee
because, based on the argument corresponding to eq.(\ref{Pj=0}), we can set
$\mathbf{P}^{(j)}$ for $j\geq 1$. Moreover
\be
\mathbf{P}^{(0)}\approx i_g \A^{(1)}\label{P0}
\ee
So (\ref{JnotdQ}) can be written simply
\be
\J_\xi \approx d\Q_\xi \label{JdQonshell}
\ee
with $\Q_\xi= \mathbf{q}_\xi + \xi i_g \A^{(1)}$ and $\Q_\xi$ given by (\ref{Qxiapp}).
For our later purposes we do not need to compute $\A^{(1)}$ explicitly.

\subsection{The variational tri-complex}
\label{app:tricomplex}

In order to justify the use of the $d, \gwedge{}$ complex, introduced in section 2, {\it on shell},
i.e. with the specification `modulo the equations of motion', the simplest way is probably to 
embed such complex in the so-called {\it variational complex}. This complex is defined by means of
the operators $d_H$ and $d_V$. Consider a space spanned locally by coordinates $x^1,\ldots, x^n$
and a set of fields $\phi^{(\alpha)}$ and their derivatives 
\be
\phi^{(\alpha)}_{I} = \phi^{(\alpha)}_{i_1 i_2\ldots i_k} =
\frac {\partial^k \phi^{(\alpha)}}{\partial x^{i_1}\partial x^{i_2}\ldots \partial x^{i_k}},\0
\ee
and define the generic functional derivative
\be
\partial_{\alpha}^I= \partial_{\alpha}^{i_1i_2\ldots i_k} = \frac {l_1!l_2!\ldots l_k!}{k!} \frac {\partial}{\partial \phi^{\alpha}_{(i_1 i_2\ldots i_k)}}\label{partialI}
\ee
where $l_j$ is the number of occurrences of the integer $j$ amongst $i_1i_2\ldots i_k$. Define moreover the contact 1-forms
\be
\theta_I^{\alpha} = \delta \phi^{(\alpha)}- \phi^{(\alpha)}_{Ij} dx^j\label{contact}
\ee
Then for any function or form $f$ made out of a
set of fields $\phi^{(\alpha)}$ and their derivatives, we have
\be
d_H f= \left( \frac {\partial f}{\partial x^i}+(\partial_{\alpha} f) \phi^{(\alpha)}_i +
(\partial^j_{\alpha}f) \phi^{(\alpha)}_{ij} +\ldots\right)dx^i\label{dH}
\ee
and
\be
d_V =  \sum_{|I|=0}^k (\partial^I_{\alpha}f) \theta^{\alpha}_I\label{dV}
\ee
$d_H$ is the form the de Rham exterior operator takes in the context of local filed theory. $d_V$ is the mathematically consistent way to represent field variations.
 
Then we have
\be
d_H^2=0,\quad\quad d_V^2=0,\quad\quad d_Hd_V +d_Vd_H=0\label{anticomm1}
\ee
These two operators define a bi-complex  by acting on the space of forms $\Omega_p^r$ of order
$p$ in space and of order $r$ in the vertical direction (that is, proportional to the exterior product of $r$ field variations). This is known as the variational bi-complex, \cite{anderson}. 

Now, it is easy to show
that
\be 
d_H \gwedge{} +\gwedge{}d_H =0, \quad\quad d_V \gwedge{} +\gwedge{}d_V =0\label{anticomm2}
\ee
thanks to the fact that the operation $\gwedge{}$ corresponds to inserting a constant Kronecker delta. We can enlarge the bicomplex to include $\gwedge{}$ and form a commutative 
{\it variational tri-complex}. $d_H,d_V$ and $\gwedge{}$ act on the spaces of forms
$\Omega_p^{(r,j)}$, where $j$ is the same as in section 2.
Here we reproduce only the relevant (for us) corner of the complex

$$\xy\xymatrix{ \ldots\ldots\!\!&\Omega_{n-2}^{(1,2)}\ar[rr]^{d_H}&& \Omega_{n-1}^{(1,2)}\ar[rr]^{d_H} && \Omega_{n}^{(1,2)}\ar[rr]^{J}&&
{\cal F}^{(1,2)}\ar[r] &0\\
&\ldots\ldots\!\!&\Omega_{n-2}^{(0,2)}\ar[ul]^{d_V}\ar[rr]^{d_H}&&\Omega_{n-1}^{(0,2)}\ar[ul]^{d_V}\ar[rr]^{d_H}&&\Omega_{n}^{(0,2)}\ar[ul]^{d_V}&&&\\
&&&&&&&&\\&&&&&&&&\\
 \ldots\ldots\!\!&\Omega_{n-2}^{(1,1)}\ar@{.>}[uuuurr]_{\gwedge{}}\ar[rr]^{d_H}&& \ar@{.>}[uuuurr]_{\gwedge{}}\Omega_{n-1}^{(1,1)}\ar[rr]^{d_H} && \Omega_{n}^{(1,1)}\ar[rr]^{J}&&
{\cal F}^{(1,1)}\ar[r] &0\\
&\ldots\ldots\!\!&\Omega_{n-2}^{(0,1)}\ar[ul]^{d_V}\ar@{.>}[uuuurr]_{\gwedge{}}\ar[rr]^{d_H}&&\Omega_{n-1}^{(0,1)}\ar[ul]^{d_V}\ar@{.>}[uuuurr]_{\gwedge{}}\ar[rr]^{d_H}&&\Omega_{n}^{(0,1)}\ar[ul]^{d_V}&&&\\
&&&&&&&&\\&&&&&&&&\\
\ldots\ldots\!\!&\Omega_{n-2}^{(1,0)}\ar@{.>}[uuuurr]_{\gwedge{}}\ar[rr]^{d_H}&& \Omega_{n-1}^{(1,0)}\ar@{.>}[uuuurr]_{\gwedge{}}\ar[rr]^{d_H} && \Omega_{n}^{(1,0)}\ar[rr]^{J}&&
{\cal F}^{(1,0)}\ar[r] &0\\
&\ldots\ldots\!\!&\Omega_{n-2}^{(0,0)}\ar[ul]^{d_V}\ar@{.>}[uuuurr]_{\gwedge{}}\ar[rr]^{d_H}&&\Omega_{n-1}^{(0,0)}\ar[ul]^{d_V}\ar@{.>}[uuuurr]_{\gwedge{}}\ar[rr]^{d_H}&&\Omega_{n}^{(0,0)}\ar[ul]^{d_V}&&&
}\endxy$$
The map $J$ is the projection that maps out total differentials. The above diagram is 
commutative and the complex is (locally) exact. The exactness for $d_H$ and $d_V$ for the bi-complex is proved by constructing the corresponding homotopy operators. The homotopy operator for $\gwedge{}$ has been constructed in section 2, (\ref{gg}). Therefore the tri-complex is exact.
The spaces ${\cal F}^{1,j}$ are spaces of equations of motion. In fact the space of equations of motion splits into subspaces labeled by $j$. For instance, $\delta_{\xi} g= \xi g^{(0)} + 
\partial \xi g^{(1)}$. Therefore $\delta_{\xi} g \, \E= \xi \E^{(0)}+ \partial \xi \E^{(1)}$.
But $\E^{(0)}$ and $\E^{(1)}$ are proportional, so they vanish simultaneously on shell. This entitles 
us to split the tri-complex according to $j$ and treat the corresponding pieces separately, as we have done in subsection \ref{app:bicomplex}.

\section{Geometry of codimension-2 surfaces}
\label{ssec:surfaces}

We review here some basics of the geometric formalism of codimension-2 surfaces ($(D-2)$-dimensional surfaces embedded in $D$-dimensional space) which we apply to a special case of bifurcation surface in Section \ref{ssec:covariance}. We follow the approach reviewed in \cite{Carter:2000wv,Booth:2006bn}, and especially Section 2 of \cite{Cao:2010vj}.

On a codimension-$p$ surface $\mathcal{S}$ embedded in a $D$-dimensional Lorentzian spacetime 
$\mathcal{M}$ one can separate the metric $g_{ab}$ of $\mathcal{M}$ in normal and tangent parts
\be \label{ghq}
g_{ab} = h_{ab} + q_{ab}
\ee
where $q_{ab}$ is known as the first fundamental tensor or induced metric on $\mathcal{S}$. In our applications $\mathcal{S}$ is spacelike, so the tangential part $q_{ab}$ is Riemannian, and the normal part 
$h_{ab}$ is Lorentzian. A decomposition (\ref{ghq}) is obtained by demanding for $q^a{}_b$ and $h^a{}_b$ to be projection operators satisfying
\be
q^a{}_b \, t^b = t^a\;, \qquad q^a{}_b \, s^b = 0 \;, \qquad h^a{}_b\, t^b = 0 \;, \qquad h^a{}_b \, s^b = s^a
\ee
for an arbitrary vector $t^a$ tangential to $\mathcal{S}$, and an arbitrary vector $s^a$ normal to $\mathcal{S}$. It  follows that $q^a{}_b$ and $h^a{}_b$ satisfy the standard projector relations
\be
q^a{}_b \, q^b{}_c = q^a{}_c \;,\qquad h^a{}_b \, h^a{}_c = h^a{}_c \;,\qquad q^a{}_b \, h^b{}_c= 0
\ee
If $\ell^a$ and $n^a$ is an arbitrary pair of two (future directed) null vector fields satisfying 
$\ell_a n^a = -1$, then we can write the normal part of the metric as
\be \label{hframe}
h_{ab} = - \ell_a n_b - n_a \ell_b
\ee
One can express $h_{ab}$ in terms of the binormal $\epsilon_{ab}$ of $\mathcal{S}$. As binormal can be written as
\be \label{binormal}
\epsilon_{ab} = \ell_a n_b - n_a \ell_b
\ee
From (\ref{hframe}) and definition of $n^a$ and $\ell^a$ it follows that
\be \label{hbinorm}
h_{ab} = \epsilon_a{}^c \epsilon_{cb}
\ee
In our application in which $\mathcal{S} = \mathcal{B}$ we can take $n^a = v^a$ and $\ell^a = u^a$, where $u^a$ and $v^a$ are vector fields obtained from Kruskal-type coordinates and defined in (\ref{uvdef}) in Appendix 
\ref{app:Kruskal}. Note that generally $l^a$ and $n^a$ are not uniquely defined, and the freedom is in the rescaling 
\be \label{lnfree}
\ell^a\to f \ell^a\;, \qquad n^a\to n^a/f 
\ee
where $f=f(x)$ is an arbitrary smooth positive function.

Another important object is the second fundamental tensor $K_{ab}{}^c$ which can be expressed as
\be \label{sftdef}
K_{ab}{}^c = q_a{}^d \, q_b{}^e \, \nabla_d q_e{}^c =  - q_a{}^d \, q_b{}^e \, \nabla_d h_e{}^c
\ee
where the second equality follows from (\ref{ghq}) and $\nabla_a g_{bc} = 0$.

Of our main interest is the normal curvature tensor $\Omega_{ab}$  which can be obtained from
\be \label{Omega2}
\Omega_{ab} &=& - \frac{1}{2} \epsilon^{cd} \left( q_a^e q_b^f h_c^g h_d^h R_{efgh} 
 + K_{aed} K_b{}^e{}_c - K_{bed} K_a{}^e{}_c \right)
\nonumber \\
&=& - \frac{1}{2} q_a^e q_b^f \epsilon^{cd} R_{efcd}
 - \frac{1}{2} \epsilon^{cd} \left( K_{aed} K_b{}^e{}_c - K_{bed} K_a{}^e{}_c \right)
\ee
There is a connection $\omega_a$, the connection of the $SO(1,1)$ normal bundle, associated with the normal curvature tensor defined by
\be \label{omega1}
\omega_a \equiv - q_a^e \, n_d \, \nabla_e \ell^d
\ee
It is important to emphasize that $\boldsymbol{\omega}$, being a connection,  is \emph{pseudo} 1-form, which means that it depends on the choice of the "frame" vectors $l^a$ and $n^a$. It is easy to check that under the rescaling (\ref{lnfree}) (which is related to change of frame) it transforms as
\be \label{gaugeom}
\omega_a \to \omega_a + q_a^b \, \nabla_b \ln f
\ee
which can be viewed as a gauge transformation of an $SO(1,1)$ connection. Note that $h_{ab}$, $q_{ab}$,
$K_{ab}{}^c$ and $\Omega_{ab} $ are instead normal (not pseudo) tensors, i.e., they are invariant on the gauge transformation (\ref{gaugeom}) (because they all can be defined without using frame vectors $l^a$ and $n^a$).

It can be shown that for codimension-2 surfaces one gets
\be \label{Om2om1}
\Omega_{ab} = q_a^c \, q_b^d (\nabla_c \omega_d - \nabla_d \omega_c)
 = q_a^c \, q_b^d (d\omega)_{cd}
\ee

In the case of our interest, when $\mathcal{S}$ is bifurcation surface $\mathcal{B}$, from (\ref{ghq}) and
(\ref{hbinorm}), due to the special property $\nabla_a \epsilon_{ab} \bifeq 0$, it follows that
\be \label{nablahq0}
\nabla_a h_{bc} = 0 \;,\qquad \nabla_a q_{bc} = 0
\ee
Using this in (\ref{sftdef}) implies that the second fundamental tensor vanishes on $\mathcal{B}$
\be
K_{ab}{}^c = 0 
\ee
Using this in (\ref{Omega2}) we finally obtain
\be \label{OmRN}
\Omega_{ab} = - \frac{1}{2} q_a^e q_b^f \epsilon^{cd} R_{efcd} = q_a^e q_b^f (R_N)_{ef}
\ee
where the 2-form $\R_N$ was defined in (\ref{GamN}).

\section{Entropy formula and spin connection}
\label{app:spinconn}

The purpose of this section is to outline a derivation of a covariant formula for the entropy in a coordinate independent
way. We defer a proof, which needs the usage of the vielbein formalism, to a future paper (for a Lorentz invariant formulation in 3D, see ref.\cite{Solodukhin:2005ah}). 
In order to appreciate the following reduction procedure it is useful to review the geometrical setting imposed upon us by the problem we are studying, see \cite{KN}, vol.II. The geometry is that of an asymptotically Minkowski space time manifold  $X$ with a codimension 2 submanifold ${\mathcal{B}}$. We have $O(X)$, the bundle of orthonormal frames on $X$ with structure group
$SO(D-1,1)$ and $O({\mathcal{B}})$
the bundle of orthonormal frames on ${\mathcal{B}}$ with structure group
$SO(D-2)$. We consider also the bundle of adapted frames. An adapted frame is a complete set of orthonormal vectors which are either tangent or orthogonal to ${\mathcal{B}}$. They form a principal bundle $O(X,{\mathcal{B}})$ with structure group
$SO(1,1)\times SO(D-2)$. To complete the description we have the bundle
of normal frames $ON({\mathcal{B}})$ with structure group
$SO(1,1)$ and the embedding $i$: $O(X,{\mathcal{B}}) \stackrel{i}{\longrightarrow} O(X)$.
For convenience, let us denote by ${\mathfrak h}$ and ${\mathfrak k}$ the Lie algebras
of $SO(D-2)$ and $SO(1,1)$, respectively.

Let us come now to formula (\ref{SCSpt}). The connection $\Gam$ in it is a connection of the linear frame bundle $LX$. Every
metric connection in $LX$ is in one-to-one correspondence with a connection in $O(X)$ (see \cite{KN}, vol.I, ch. 4, $\S$ 2). So we assume we can replace (\ref{SCSpt}) with
\be \label{SCSpta}
S_{\mathrm{CS}} = 2\pi n(n-1) \int_0^1 dt \int_{\mathcal{B}} P_n(\hat\epsilon,\hat\al,\hat\R_t^{n-2})
\ee
where $\hat\al$ is the reduction of $\Gam$ to the structure group $SO(1,D-1)$ and $\hat \R$ its curvature. Moreover $\hat\epsilon= E\epsilon E^{-1}$,
where $E=\{E_\mu^a\}$ are the vielbein in $X$ ($a$ is a flat index, see below). $P_n$ is now a symmetric polynomial in the Lie algebra of $SO(D-1,1)$.

In (\ref{SCSpta}) it is understood that the forms in the integrand are pulled back from $X$
to ${\mathcal{B}}$. Now by pulling back a generic connection $\hat\al$ of $O(X)$ through $i$, we do not get a connection, unless we restrict to the components in ${\mathfrak h}$+
${\mathfrak k}$. If so, the connection splits into
$\hat\al_t+\hat\al_n$, that is a connection  $\hat\al_t$ in $O({\mathcal{B}})$ with values
in $\mathfrak h$ and a connection $\hat\al_n$ in $ON({\mathcal{B}})$ with values in $\mathfrak k$ (see \cite{KN}, vol.II, ch. VII). As we have clarified in section 4, the geometry of the problem is defined by the presence of the surface ${\mathcal{B}}$ with its tangent and normal directions, thus the just considered reduction of a connection pulled back from $X$,
is natural in this scheme. But once we replace in (\ref{SCSpta}) the connection $\hat\al_t+\hat\al_n$, with values in the direct sum ${\mathfrak h}$+ ${\mathfrak k}$, the presence
of the binormal $\epsilon$ maps out the ${\mathfrak h}$ components and only the components
along ${\mathfrak k}$ (the Lie algebra of the normal frame bundle structure group $SO(1,1)$) survive. At this point we
have to do with an Abelian connection and we can easily integrate over $t$ as in section 4.1.

To view the situation in more detail let us introduce the following conventions. Let us denote by $\mu,\nu,..=0,\ldots,D-1$ generic curved indices and $a,b,\ldots=0,\ldots,D-1$
generic flat indices. Then, following  \cite{Carter:2000wv}, we will denote
by $A,B,..=2, \ldots, D-1$ flat tangent indices in ${\mathcal{B}}$ and by $X,Y,\ldots=0,1$ normal flat indices, and introduce
adapted vielbein $\imath_A{}^\mu$ and $\lambda_X{}^\mu$ (they are particular cases of $E_a^\mu$) , such that
\be
q^{\mu}{}_\nu = \imath_A{}^\mu \imath^A{}_\nu,\quad\quad h^\mu{}_\nu = \lambda_X{}^\mu \lambda^X{}_\nu\label{qandh}
\ee
One can show in particular that, since $ \lambda_X{}^\mu\lambda_{Y\mu} =\eta_{XY}$ ($\eta$ denotes the flat Minkowski metric), one can make the following identifications 
\be
\lambda_0{}^\mu= \frac {n^\mu-l^\mu}{\sqrt{2}},\quad\quad \lambda_1{}^\mu= \frac {n^\mu+l^\mu}{\sqrt{2}}\label{lambdanl}
\ee 
with reference to the null vectors introduced in the previous Appendix.
Then, it is easy to show that
\be
\epsilon_{\mu\nu} E_a^\mu E_b^\nu=\eta_{1a} \eta_{0b}-\eta_{1b} \eta_{0a} \label{eEE}
\ee 
where $\eta$ is the flat Minkowski metric. Thus, for instance,
\be
\tr (\epsilon \hat\al_n)=\hat\epsilon_{\mu\nu} E_a^\mu E_b^\nu (\hat\al_n)^{ab}= 2 \hat\al_n^{01}\label{alpha01}
\ee
and likewise for the curvature.  Therefore in this approach we obtain formulas like (\ref{GamN}) and  (\ref{ourTach}) with
$\Gam$ and $\R$ replaced by $\hat\al_n$ and its curvature $\hat \R_n$. It is understood that all the forms 
are pulled back to ${\mathcal{B}}$, which can be achieved on components by contracting the form index with the $q$ 
projector: for instance the intrinsic component of the pulled back $\hat\al_n$ is $q_\mu{}^\nu (\hat\al_n)_\nu$.

It is now convenient to compare the normal bundle connection  with the one introduced in \cite{Carter:2000wv},
\be
\varpi_\mu{}^\nu{}_\rho = h_\sigma{}^\nu \lambda^X{}_\rho \bar \nabla_\mu \lambda_X{}^\sigma, \quad {\rm where}\quad 
\bar \nabla_\mu= q^\nu{}_\mu \nabla_\nu\label{varomega}
\ee
Using $\nabla E_a^\mu= -\hat\al_a{}^b E_b^\mu$ we can rewrite
\be
\varpi_\mu{}^\nu{}_\rho= q_\mu{}^\sigma \epsilon_\rho{}^\nu (\hat\al_n)_\sigma^{01}\label{varpi2}
\ee
Saturating with $\epsilon_\nu{}^\rho$ we obtain precisely the RHS of (\ref{alpha01}).

On the other hand, inserting (\ref{lambdanl}) into (\ref{varomega}) one finds
\be 
\varpi_\mu{}^\nu{}_\rho = -\epsilon^\nu{}_\rho n_\tau \bar \nabla_\mu \ell^\tau \label{varpi3}
\ee
Saturating with $\epsilon^\rho{}_\nu$ and dividing by 2, we get precisely the definition (\ref{omega1}).

Finally a comment about gauge transformations in the normal frame bundle. They are valued in $SO(1,1)$ and act on $l_0,l_1$ as follows
\be
\left(\begin{matrix} l_0 \\ l_1\end{matrix} \right) \rightarrow \left(\begin{matrix} \cosh t& \sinh t \\ \sinh t& \cosh t\end{matrix} \right)\left(\begin{matrix} l_0 \\ l_1\end{matrix} \right)\label{so11}
\ee
where $t$ is a local function. Using again (\ref{lambdanl}), it is easy to see that they act on $n,l$ as a rescaling
\be
n\rightarrow e^t n,\quad\quad l\rightarrow e^{-t} l\label{transfnl}
\ee
Identifying $1/f=e^t$ we get precisely the rescalings considered in (\ref{gaugeom}).

\section{Kruskal-type coordinates}
\label{app:Kruskal}

Here we prove some of the relations and properties used in Section  \ref{sec:CSentropy}. The strategy
we use is to first make calculations in special ``Kruskal-type" coordinates, and then eventually generalizing to other coordinate systems typically used in black hole calculations. 

In \cite{Racz:1992bp} it was shown that, in a spacetime with Killing horizon on which the surface gravity is constant, one can construct Kruskal-type coordinates $(U,V,\{x^i\})$, $i=1,\ldots,D-2$, in which metric has the following form
\be \label{kruskal}
ds^2 = G\, dU dV + V H_i\, dx^i dU + g_{ij}\, dx^i dx^j
\ee
where $G$, $H_i$ and $g_{ij}$ are generally smooth functions of $D-1$ variables $U,V$ and $\{x^i\}$. 
The physical horizon is at $U=0$, while $U=V=0$ defines the bifurcation surface $\mathcal{B}$. 
On the bifurcation surface $\mathcal{B}$ we have $G|_{\mathcal{B}} = -2/\kappa$ 
where $\kappa$ is the surface gravity and is constant throughout $\mathcal{B}$. 
We see that 
$\{x^i\}$ are tangential and $U,V$ are normal on $\mathcal{B}$.  The horizon generating Killing vector field 
$\xi$ is given by
\be \label{xiKrus}
\xi = \kappa \left( U \frac{\dd}{\dd U} - V \frac{\dd}{\dd V} \right)
\ee
where the constant $\kappa$ is surface gravity. In Kruskal coordinates, the components of the metric (\ref{kruskal}) are regular and well-defined on $\mathcal{B}$, and the components of $\xi$ obviously satisfy
\be
\xi^a \big|_\mathcal{B} = 0 \quad , \qquad
 \nabla_b \xi^a \big|_\mathcal{B} = \partial_b \xi^a \equiv \Lambda^a_{\;\; b} 
\ee
and the nonvanishing components of $\Lambda^a_{\;\; b}$ are
\be \label{lambdaKrus}
\Lambda^U_{\;\; U} = - \Lambda^V_{\;\; V} = \kappa
\ee
From this it follows that
\be \label{dLKrus}
d \Lambda^a_{\;\; b} \equiv \partial_\mu \Lambda^a_{\;\; b} dx^\mu  = 0
\ee

Using Kruskal-type coordinates one can introduce a pair of null vector fields $U^\mu$ and $V^\mu$
normal to $\mathcal{B}$ defined by
\be
U^\mu = \left( \frac{\dd}{\dd U} \right)^\mu \;, \qquad
V^\mu = \left( \frac{\dd}{\dd V} \right)^\mu \;, \qquad
\ee
If we make a rescaling
\be \label{uvdef}
u^\mu \equiv \frac{U^\mu}{\sqrt{-U_\mu V^\mu}} \;, \qquad v^\mu \equiv \frac{V^\mu}{\sqrt{-U_\mu V^\mu}}
\ee
then $U^\mu$ and $V^\mu$ are a pair of future-oriented null vector fields normal to $\mathcal{B}$
which is normalized such that
\be \label{uvnorm}
u_\mu \, v^\mu = -1
\ee

Using Kruskal coordinates we can obtain important information on the components of connection and 
Riemann tensor on the bifurcation surface $\mathcal{B}$. We start with the Killing lemma
\be\label{Killemma}
\nabla_{\!a} \nabla_{\!b} \, \xi^c = \tensor{R}{^c_b_a_d} \, \xi^d
\ee
satisfied by any Killing vector field. On $\mathcal{B}$ the left hand side of (\ref{Killemma}) can be written as
\be
\nabla_{\!a} \nabla_{\!b} \, \xi^c \, \bifeq \, \Gamma^c_{bp} \, \tensor{\Lambda}{^p_a} - \Gamma^r_{ab} \, \tensor{\Lambda}{^c_r} + \Gamma^c_{as} \, \tensor{\Lambda}{^s_b}
\ee
so that,
\be\label{Gammas}
\Gamma^c_{bp} \, \tensor{\Lambda}{^p_a} - \Gamma^r_{ab} \, \tensor{\Lambda}{^c_r} 
+ \Gamma^c_{as} \, \tensor{\Lambda}{^s_b} \, \bifeq \, 0
\ee
By making particular choices for free indices in (\ref{Gammas}) we get
\be
\Gamma^U_{ij} = \Gamma^V_{ij} = \Gamma^U_{Vi} = \Gamma^V_{Ui} = \Gamma^i_{jU}
 = \Gamma^i_{jV} = \Gamma^i_{UU} = \Gamma^i_{VV} = 0
\ee
and
\be
\Gamma^U_{UU} = \Gamma^V_{VV} = \Gamma^U_{VV} = \Gamma^V_{UU} = \Gamma^U_{UV} = \Gamma^V_{UV} = 0
\ee
All this equalities should be understood as $\bifeq$, which, for brevity, we avoid to write here. It
follows that the only nonvanishing components of the pull-back of the tensor valued 1-form connection 
$\mathbf{\Gamma}$ on the bifurcation surface $\mathcal{B}$ are $\Gamma^U_{Ui}$, $\Gamma^V_{Vi}$ and $\Gamma^k_{ji}$. One important consequence is
\be \label{gam2bs}
\Gamma^a_{ci}\, \Gamma^c_{bj} \, dx^i \wedge dx^j \, \bifeq \, 0 \;,\quad \mbox{when } a \mbox{ and/or }
 b \mbox{ is } U \mbox{ or } V \;.
\ee

By taking the covariant derivative of (\ref{Killemma}) we get
\be
\nabla_{\!p} \nabla_{\!a} \nabla_{\!b} \, \xi^c \, \bifeq \, \tensor{R}{^c_b_a_d} \, \tensor{\Lambda}{^d_p} = \left( \Gamma^c_{bd,a} - \Gamma^c_{ba,d} + \Gamma^c_{ae} \, \Gamma^e_{bd} - \Gamma^c_{df} \, \Gamma^f_{ba} \right) \tensor{\Lambda}{^d_p}
\ee
while from the definition of covariant derivative we have
\be
\nabla_{\!p} \nabla_{\!a} \nabla_{\!b} \, \xi^c &\bifeq& \Gamma^c_{bd,a} \, \tensor{\Lambda}{^d_p} + \Gamma^c_{ae,p} \, \tensor{\Lambda}{^e_b} + \Gamma^c_{bf,p} \, \tensor{\Lambda}{^f_a} - \Gamma^g_{ab,p} \, \tensor{\Lambda}{^c_g} + \nonumber\\
&& + \left( \Gamma^c_{ar} \, \Gamma^r_{bh} - \Gamma^s_{ab} \, \Gamma^c_{sh} \right) \tensor{\Lambda}{^h_p}
\ee
Equating right hand sides gives us
\be\label{dGammas}
\Gamma^c_{ba,d} \, \tensor{\Lambda}{^d_p} + \Gamma^c_{ae,p} \, \tensor{\Lambda}{^e_b} + \Gamma^c_{bf,p} \, \tensor{\Lambda}{^f_a} - \Gamma^g_{ab,p} \, \tensor{\Lambda}{^c_g} \, \bifeq \, 0
\ee
Furthermore, from (\ref{dGammas}) we have
\be
\Gamma^U_{ij,k} = \Gamma^V_{ij,k} = \Gamma^U_{Vi,j} = \Gamma^V_{Ui,j} = \Gamma^i_{Uj,k} = \Gamma^i_{Vj,k} = 0
\ee
Again, all the above equations should be understood as $\bifeq\,$.  

Now we focus on the Riemann tensor $\tensor{R}{^a_b_\mu_\nu}$. As we do the integration over the bifurcation surface, the only relevant components are those for which the ``form indices'' $\mu$ and 
$\nu$ relate to coordinates from the subset $\{x^i\}$. Using results from above inside the the definition of Riemann tensor,
\be
\tensor{R}{^a_b_\mu_\nu} = \Gamma^a_{b\nu,\mu} - \Gamma^a_{b\mu,\nu} + \Gamma^a_{\mu\sigma} \Gamma^\sigma_{b\nu} - \Gamma^a_{\nu\rho} \Gamma^\rho_{b\mu}
\ee
we obtain
\be \label{Riembs}
\tensor{R}{^U_k_i_j} \,\bifeq\, 0 \qqd \tensor{R}{^k_U_i_j} \,\bifeq\, 0 \qqd \tensor{R}{^V_k_i_j} \,\bifeq\, 0 \qqd \tensor{R}{^k_V_i_j} \,\bifeq\, 0
\ee
In other words, the 2-form $\tensor{R}{^a_b_i_j}$ is block-diagonal in the tensor indices when pulled-back on the bifurcation surface $\mathcal{B}$.

We can write the obtained properties of $\nabla_b \xi^a$ and Riemann tensor in the following covariant way
\begin{eqnarray}
&& 0 \, \bifeq \, q^c{}_b \, \nabla_c \xi^a \, \bifeq \, q^a{}_c \nabla_b \xi^c \; \Longrightarrow \;
 0 \, \bifeq \, q^c{}_b \, \epsilon{}^a{}_c \, \bifeq \, q^a{}_c \epsilon{}^c{}_b
\label{nabxicov} \\
&& 0 \, \bifeq \, h^a{}_e \, q^f{}_b \, q^g{}_c \, q^r{}_d \, R^e{}_{fgr} \, \bifeq \, 
q^a{}_e \, h^f{}_b \, q^g{}_c \, q^r{}_d \, R^e{}_{fgr}
 \label{Riebscov}
\end{eqnarray}
where $q^a{}_b$ and $h^a{}_b$ are tensors obtained by separation of metric tensor $g^a{}_b$ into
tangent and normal part to $\mathcal{B}$, defined in Appendix \ref{ssec:surfaces}. Being written in a covariant way, relations in (\ref{nabxicov}) and (\ref{Riebscov}) are valid in \emph{all} coordinate 
systems.

\section{Schwarzschild-type coordinates for stationary black holes}
\label{app:Sch}

Here we give the proof of (\ref{GNSK}) in the case of stationary geometry. The relation between Kruskal-type coordinates $(U,V,\{x^i\})$ (described in Appendix \ref{app:Kruskal}) and Schwarzschild-type coordinates 
$(t,r,\{x^{i'}\})$ was constructed in \cite{Racz:1992bp}, and is given by
\be \label{cset1}
t &=& \frac{1}{2\kappa} \ln \left[ -\frac{U}{V} \exp\left(2\kappa H(UV,\{x^i\})\right) \right] \0 \\
r &=& UV f(UV,\{x^i\}) \label{KrSch} \\
x^{i'} &=& x^i \;, \quad i = 2,\ldots, D-2 \quad \mbox{(coordinates tangent to $\mathcal{B}$)} \0
\ee
where $f$ and $H$ are some smooth functions around $\mathcal{B}$ (defined by $U=V=0)$, which
are non-vanishing on $\mathcal{B}$. 

The difference between $\Gam_N$ evaluated in two (arbitrary) coordinate systems $\{x^a\}$ and 
$\{x^{a'}\}$ is
\be
\hat{\Delta} \Gam_N = \frac{1}{2} \tr(\epsilon d\Lambda \Lambda^{-1})
 = - \frac{1}{2} \tr(\epsilon \Lambda d\Lambda^{-1}) 
\ee
where $(\Lambda){}^a{}_b \equiv \partial_{b'} x^{a}$ and $(\Lambda^{-1}){}^a{}_b = \partial_{b} x^{a'}$.

When we specialize to $\{x^a\} = (U,V,\{x^i\})$ (identified below by index $K$, meaning Kruskal-type) we obtain
\be \label{delKGN}
2 \hat{\Delta}_K \Gam_N = (\Lambda_K d\Lambda_K){}^V{}_V - (\Lambda_K d\Lambda_K){}^U{}_U
\ee
There is no reason for the expression in (\ref{delKGN}) to be vanishing in general. However, we shall now show that this happens when we take $\{x^{a'}\} = (t,r,\{x^i\})$, i.e., Schwarzschild-type. 
First of all, note that integration over $\mathcal{B}$ forces form-indices to take values $i$, $i=2,\ldots,D-1$, 
so we need to calculate
\be \label{delKGNi}
2 \hat{\Delta}_K (\Gam_N)_i 
 = \frac{\partial V}{\partial x^{a'}}\frac{\partial^2 x^{a'}}{\partial V \partial x^i}
  - \frac{\partial U}{\partial x^{a'}}\frac{\partial^2 x^{a'}}{\partial U \partial x^i}
\ee
From (\ref{KrSch}) it follows that the nonvanishing second derivatives present in (\ref{delKGNi}) 
are\footnote{As already commented before, when calculating in singular coordinate systems we have 
to avoid using $U=V=0$ in the middle of calculations. This is why we are calculating for finite 
$U,V$, and only at the end of the calculation we take $U,V \to 0$.}
\be
&&\frac{\partial^2 t}{\partial U \partial x^i} = V \partial_i \dot{H} \;, \qquad
\frac{\partial^2 t}{\partial V \partial x^i} = U \partial_i \dot{H} \;, \0 \\
&&\frac{\partial^2 r}{\partial U \partial x^i} = V \left( \partial_i f + UV \partial_i \dot{f} \right) \;, \quad
\frac{\partial^2 r}{\partial V \partial x^i} = U \left( \partial_i f + UV \partial_i \dot{f} \right) \0
\ee
where a dot on a function means $\dot{g} \equiv \partial_{UV} g$. We also need
\be
\frac{\partial U}{\partial t} = \kappa U \;, \quad 
\frac{\partial U}{\partial r} = \frac{1}{2V f(0)} + \mathcal{O}(U) \;, \quad
\frac{\partial V}{\partial t} = - \kappa V \;, \quad
\frac{\partial V}{\partial r} = \frac{1}{2U f(0)} + \mathcal{O}(V)
\ee
where $f(0) \equiv f(0,\{x^i\})$. By using all this in (\ref{delKGNi}) we obtain that the lowest order terms in 
(\ref{delKGNi}) cancel and
\be
\hat{\Delta}_K (\Gam_N)_i = \mathcal{O}(UV) \stackrel{\mathcal{B}}{\longrightarrow} 0
\ee
which implies (\ref{GNSK}).

Note that in coordinates $(t,r,\{x^{i'}\})$, the horizon generating Killing vector $\xi$ is equal to $\partial/\partial t$. 
This means that (\ref{cset1}) does not include coordinates of the Boyer-Lindquist type where $\xi$ is of the form 
$\xi = \partial/\partial t + \Omega \partial/\partial \phi$. Fortunately, it is straightforward to extend the result 
$\left. \hat{\Delta}_K (\Gam_N)_i \right|_\mathcal{B} = 0$ to the following more general transformations which include coordinates of the Boyer-Lindquist type:
\be \label{cset2}
t &=& \frac{1}{2\kappa} \ln \left[ -\frac{U}{V} \exp\left(2\kappa H(UV,\{x^i\})\right) \right] \0 \\
r &=& UV f(UV,\{x^i\}) \0 \\
{x}^{i'} &=& {x}^{i'}(U,V,\{x^j\}) \;, \quad i,j = 2,\ldots, D-1 \quad \mbox{($x^j$ are coordinates tangent to $\mathcal{B}$)} \0
\ee
where the Jacobian $\left|\frac{\partial {x}^{a'}}{\partial {x}^b}\right| \neq 0$ is well defined on $\mathcal{B}$ ($a,b=0,\ldots,D-1$).


\end{document}